\documentclass[a4paper,prb,twocolumn,showpacs]{revtex4}
\usepackage{amsfonts}
\usepackage{ifthen}
\usepackage{ulem}

\usepackage[pdftex]{graphicx}\pdfcompresslevel=9
\usepackage[pdftex,breaklinks=true,colorlinks=true,]{hyperref}
	
\newcommand{\La}{\line (1,0  ){12}}
\newcommand{\Lb}{\line (3,5 ){6}}

\newcommand{\Ld}{\line (-1,0){12}}
\newcommand{\Le}{\line (-3,-5){6}}

\newcommand{\C} {\circle*{4}}

\newcommand{\pA}{\put(-6,-10)}
\newcommand{\pB}{\put(6,-10)}
\newcommand{\pC}{\put(12,0)}

\newcommand{\pZ}{\put(0,0)}

\newcommand{\rhomb}{
  \pA{\C}\pB{\C}\pZ{\C}\pC{\C}
 }

\newcommand{\rhombH}{
  \begin{picture}(22,10)(-8,-6)
    \rhomb
    \pA{\La}\pC{\Ld}
  \end{picture}
}
\newcommand{\rhombV}{
  \begin{picture}(22,10)(-8,-6)
    \rhomb
    \pB{\Lb}\pZ{\Le}
  \end{picture}
}

\newcommand{\Z}{\mathbb{Z}_2}

\newcommand{\journal}[4]{\ifthenelse{\equal{#1}{prl}}{\href{http://link.aps.org/abstract/PRL/v#2/e#3}{\prl {\bf #2}, #3 (#4)}}{\ifthenelse{\equal{#1}{prb}}{\href{http://link.aps.org/abstract/PRB/v#2/e#3}{\prb {\bf #2}, #3 (#4)}}{\ifthenelse{\equal{#1}{arxiv}}{\href{http://arxiv.org/abs/#2.#3}{arXiv:#2.#3}}{\ifthenelse{\equal{#1}{rmp}}{\href{http://link.aps.org/abstract/RMP/v#2/e#3}{\rmp {\bf #2}, #3 (#4)}}{\ifthenelse{\equal{#1}{cond-mat}}{\href{http://arxiv.org/abs/cond-mat/#2}{cond-mat/#2}}{{#1 {\bf #2}, #3 (#4)}}}}}}}

\begin{document}

\title{
Quantum Dimer Model on the triangular lattice:
Semiclassical and variational approaches to vison dispersion and condensation
}

\author{Gr\'egoire Misguich}
\email{gregoire.misguich@cea.fr}
\affiliation{Institut de Physique Th\'eorique, CEA Saclay, 91191 Gif-sur-Yvette Cedex, France}
\author{Fr\'ed\'eric Mila}
\affiliation{Institute of Theoretical Physics, Ecole Polytechnique F\'ed\'erale de Lausanne, CH-1015 Lausanne, Switzerland}

\begin{abstract}
 After reviewing the concept of vison excitations in $\Z$ dimer liquids, we
 study the liquid-crystal transition of the Quantum Dimer Model
 on the triangular lattice by means of a semiclassical spin-wave
 approximation to the dispersion of visons in the context of a ``soft-dimer" 
version of the model.
 This approach captures some important qualitative
 features of the transition: continuous nature of the transition,
 linear dispersion at the critical point, and $\sqrt{12}\times\sqrt{12}$
symmetry-breaking pattern. In a second part, we present a variational calculation
of the vison dispersion relation at the RK point which
reproduces the qualitative shape of the dispersion relation and
the order of magnitude of the gap. This approach provides
a simple but reliable approximation of the vison wave functions
at the RK point.
\end{abstract}

\date{January 23rd, 2008}


\pacs{05.50.+q,71.10.-w,75.10.Jm}

\maketitle

\section{Introduction}

Since they have been shown to possess Resonating Valence Bond (RVB) phases
on the triangular,\cite{ms01b} kagome\cite{msp02}
and other (non-bipartite) lattices,\cite{ms03}
Quantum Dimer Models (QDM) have been one of the main
paradigms in the field of quantum spin liquids.
These models, where the Hilbert space is spanned by hard-core dimer coverings of the lattice,
are expected to capture the phenomenology of quantum antiferromagnets where the wave function
is dominated by short-range valence bond configurations.
On the triangular
lattice, the
simplest
QDM is defined by the Hamiltonian:
\begin{eqnarray}
  \mathcal{H}=&-t&\sum_r \left(
  \left|\rhombV\right>\left<\rhombH\right| +{\rm H.c.}
  \right) \nonumber \\
  &+V&\sum_r\left(
  \left|\rhombV\right>\left<\rhombV\right|+\left|\rhombH\right>\left<\rhombH\right|
  \right)
\label{eq:tQDM}
\end{eqnarray}
where the sum runs over all plaquettes (rhombi) including the three possible 
orientations. The kinetic term,  controlled by $t$, flips the
two dimers on every flippable plaquette, i.e., on every plaquette with two
parallel dimers, while the potential term controlled 
by the interaction $V$ describes a repulsion ($V>0$) or an attraction ($V<0$) 
between nearest-neighbor dimers.

The RVB phase is
now
relatively well understood. The first result goes back to
Rokhsar and Kivelson\cite{rk88} who showed that, for $V/t=1$ (the Rokhsar-Kivelson or RK point), 
the ground state is the sum of all configurations with equal amplitudes:
\begin{equation}
 | RK\rangle = \frac{1}{\sqrt{\mathcal N}} \sum_c| c \rangle.
\end{equation}
Since then, dimer-dimer correlations have been shown to be short-ranged in a range
of parameters below $V/t=1$,
\cite{ms01b,ioselevich02} and the excitation spectrum to be 
gapped.\cite{ms01b,ivanov04,ralko05,lca07} Besides, it has topologically degenerate 
ground states on non-simply connected clusters.\cite{ms01b,ioselevich02,ralko05} 
This degeneracy is not related to a standard symmetry breaking. Indeed, there is no
local order parameter,\cite{ioselevich02,fmo06}
but only $\Z$ topological order.\cite{fm07}

In QDM, the nature of the phase transition 
from a liquid to a solid
is a long  standing problem.
It goes back to Jalabert and 
Sachdev\cite{js91}
(see also \cite{sv00})
who studied a
 three-dimensional frustrated Ising model
 related to the square lattice QDM.

In a more general context, Senthil and Fisher\cite{sf00,sf01}
showed that models of Mott insulators can
be cast in the form of a $\Z$ gauge theory.
In this language, the transition from a fractionalized insulator to
a conventionally ordered insulator appears to be a
condensation of $\Z$-vortices (dubbed {\it visons}).
Using a duality relation,\cite{wegner71}
they showed that such transitions correspond to an ordering transition in a {\it frustrated} Ising model
in transverse field.
As we will see, this applies to the present QDM.

Building on a mapping between QDM's at $V=0$ and Ising models in 
a transverse field,
Moessner
{\it et al.} \cite{msc00,ms01}
have developed a Landau-Ginzburg approach
and
suggested that the transition could be continuous
and in the three-dimensional $O(4)$ universality class.

Numerical evidence in favor of this scenario has been obtained
with Green's function Quantum Monte Carlo by Ralko {\it et al.}, who have shown that,
at the transition point, the static form factor of the crystal decreases
to zero on the crystal side of the transition, while the dimer gap 
also decreases to zero on the liquid side.\cite{ralko06} More recently, the vison
spectrum has also been numerically determined using Green's function Quantum Monte Carlo,
with the conclusion that a soft mode indeed develops at the
transition.\cite{ralko07}

In spite of these  results,
a simple picture for the wave functions of the visons
and the evolution of their spectrum 
is still missing.
One difficulty is
that, like any vortex, these excitations cannot be created by operators which are local (in the dimer variables).
To attack this
problem, we follow two strategies. First of all, we look at the problem in
the context of a $\Z$ gauge theory on the triangular lattice. As usual, this theory can be 
mapped onto a dual Ising model.\cite{wegner71} The duality transforms the nonlocal excitations of the 
gauge theory into local excitations, which we study using a semiclassical approach.
This simple $1/S$ expansion already captures most aspects of the 
confinement-deconfinement transition (soft mode and condensation) of the $\Z$ gauge theory.
Secondly, building on the explicit form of the excitations of the $\Z$ gauge theory,
we construct single vison wave functions for the QDM and study the properties
of the system at the RK point in the variational Hilbert space spanned by these states.
These ``variational visons'' -- living on the triangular {\it plaquettes}
and experiencing a flux $\pi$ emanating from each {\it site} of the lattice --
turn out to be linearly independent.
The associated  dispersion is in good qualitative agreement with that obtained in Monte Carlo simulations
and the value of the gap at the RK point ($\Delta_{\rm var}=0.119$), obtained with a single variational parameter,  has the correct order of magnitude (Monte Carlo simulations\cite{ivanov04} give 0.089).
Improving quantitatively further these variational results would require more adjustable parameters to account
in more details for the local (dimer-dimer, etc) correlations in the vicinity of the core of the vortex,
a task which has not been carried out here.

\section{Visons in $\Z$ gauge theory}

The connection between QDMs and $\Z$ gauge theories has already
been discussed 
from different perspectives (see in particular Refs.~\onlinecite{sv00,msf02,msp02,ms03}),
and is rooted in the existence, in both families of models, of Ising-like degrees of freedom subjected to local constraints (hard-core constraints for the dimers, Gauss law for the gauge theory).

In this section, 
we review two known mappings:
i) from a $\Z$ lattice gauge theory
on the hexagonal lattice to the triangular lattice QDM
(valid in a particular limit, Sec.~\ref{ssec:Z2}) and ii) from the $\Z$ 
theory to its dual frustrated Ising model (Sec.\ref{ssec:dim}).
The latter Ising model is then studied using a semiclassical approximation
in Secs.~\ref{sssec:cpd} and \ref{sssec:sca}.
Since the $\Z$ gauge theory - QDM mapping is formally justified
only in the confined  phase of the
gauge theory, the relevance of the results obtained in the RVB (deconfined) phase of the 
QDM is not guaranteed {\it a priori}. Some reasons why this is expected to
be the case are discussed in the next section, when we build on the results obtained
for the excitations of the $\Z$ gauge theory in its deconfined phase to construct elementary excitations
in the RVB phase of the QDM.

\subsection{$\Z$ gauge theory}
\label{ssec:Z2}

\begin{figure}
 \includegraphics[height=3.5cm]{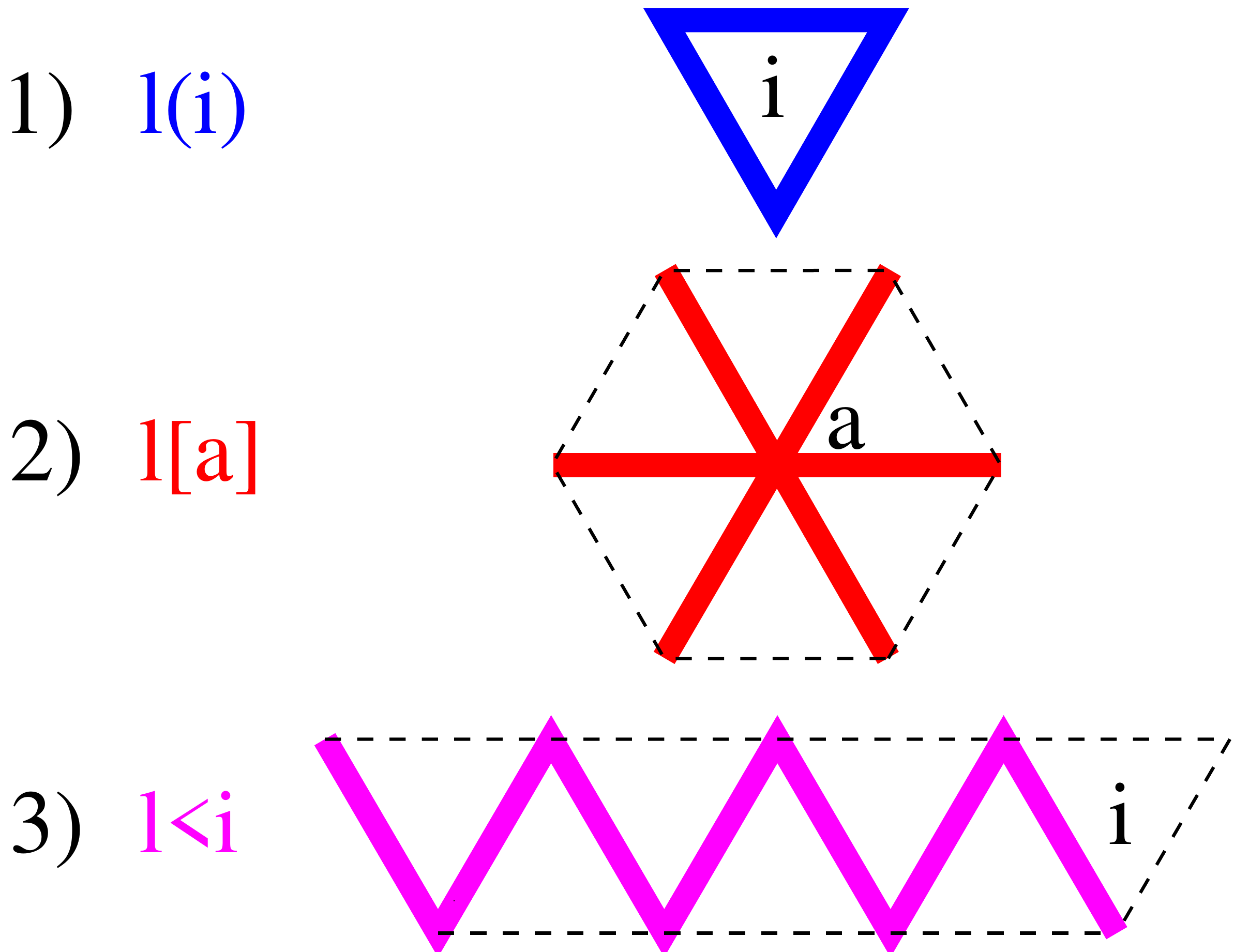}
 \caption{(Color online) Some useful definitions of sets of bonds:
 1) $l(i)$ is the set of the three bonds forming the edges of the plaquette
 $i$.
 2) $l[a]$ is the set of the six (fat) bonds emanating from the triangular site $a$.
 3) $l<i$ denotes the bonds forming a ``zigzag'' string extending (to the left) from the triangular
 plaquette $i$ to an edge of the lattice.}
 \label{fig:def}
\end{figure}

To write down a $\Z$ gauge theory analogous to a QDM, 
the starting point is to define Pauli matrices $\vec \tau_l$ on the 
bonds $l$ of the triangular lattice such that $\tau^x_l=-1$ if bond $l$ 
is occupied by a dimer and $+1$ if it is empty, while $\tau^z_l$ changes
the state of bond $l$. Since each kinetic term of the QDM flips the dimers
around a rhombus, it corresponds to a  
product of four $\tau^z_l$ around this rhombus. This would, however, not keep  the structure of the simple
$\Z$ 
gauge theories, in which the kinetic term acts on an elementary plaquette of the lattice,
a crucial ingredient to get a simple Ising model by duality.\cite{wegner71} An alternative is to consider
the more standard $\Z$ gauge theory which, in its Hamiltonian formulation, 
is defined by:
\begin{equation}
H=H_J+H_\Gamma=- J \sum_l \tau^x_l  - \Gamma \sum_i
\prod_{l(i)} \tau^z_{l(i)};
\end{equation}
where $i$ runs over the sites of the dual honeycomb lattice, and $l(i)$ are the three 
bonds forming the triangular plaquette around site $i$ (see Fig.~\ref{fig:def}). The hallmark of this model
is to have local conserved quantities. Indeed, 
\begin{equation}
[H,\prod_{l[a]} \tau^x_{l[a]}]=0,
\end{equation}
where $a$ is a site of the triangular lattice, and the product over $l[a]$ runs
over the six links emanating from $a$ (see Fig.~\ref{fig:def}). This allows one to define different sectors according
to whether $\prod_{l[a]} \tau^x_{[a]}$ is equal to $+1$ or $-1$. 
Since in the QDM the number of dimers emanating from a given site is exactly equal
to 1, it is clearly better to consider the sector where
\begin{equation}
\prod_{l[a]} \tau^x_{l[a]}=-1
\end{equation}
for all $a$ since this forces the number of dimers emanating from a site to be odd
(defining an {\it odd Ising gauge theory} in the terminology of Ref.~\onlinecite{msf02}).
Then, the true constraint is recovered in the limit $\Gamma/J\rightarrow 0$
if $J>0$ since, in the ground state, the number of dimers is then minimal. 
A {\it bona fide} QDM is then recovered if $H_\Gamma$ is treated with degenerate perturbation
theory. Since $H_\Gamma$ changes the number of dimers, its effect vanishes to first order.
To second order however, one recovers exactly the QDM of Eq. (1) with $V=0$ and $t=\Gamma^2/J$.
So, if $J/\Gamma\gg 1$, the $\Z$ gauge theory maps onto the QDM at $V/t=0$.
At finite $J/\Gamma$ the model can be viewed as a ``soft-dimer" model, where 1, 3 or 5 dimers
may touch a given site.

As we shall see below, the limit $J/\Gamma\gg 1$ lies deep inside the confined phase of the gauge
theory, and from previous work on the QDM, it is known that at $V/t=0$, the model is in a
Valence Bond Crystal phase.
It would, of course, be very interesting to connect the two models
away from this limit, when the gauge
theory is in its deconfined phase and the QDM is in the dimer liquid phase.
An interesting step in this direction is provided by higher order
perturbation theory. Indeed, to fourth order in $\Gamma/J$, a repulsion between dimers 
$V=\Gamma^4/2J^3$ is generated. However, other terms of the same order,
involving dimer shifts
along loops of length 6, are also generated (see Appendix \ref{sec:4th}). So, a {\it rigorous}
mapping does not extend beyond the $V/t=0$ point.

\subsection{Dual Ising model}
\label{ssec:dim}

As usual, this $\Z$ gauge theory is best analyzed by mapping it onto a dual Ising model 
in a transverse field.

\subsubsection{The model}

\begin{figure}
 \includegraphics[height=3.8cm]{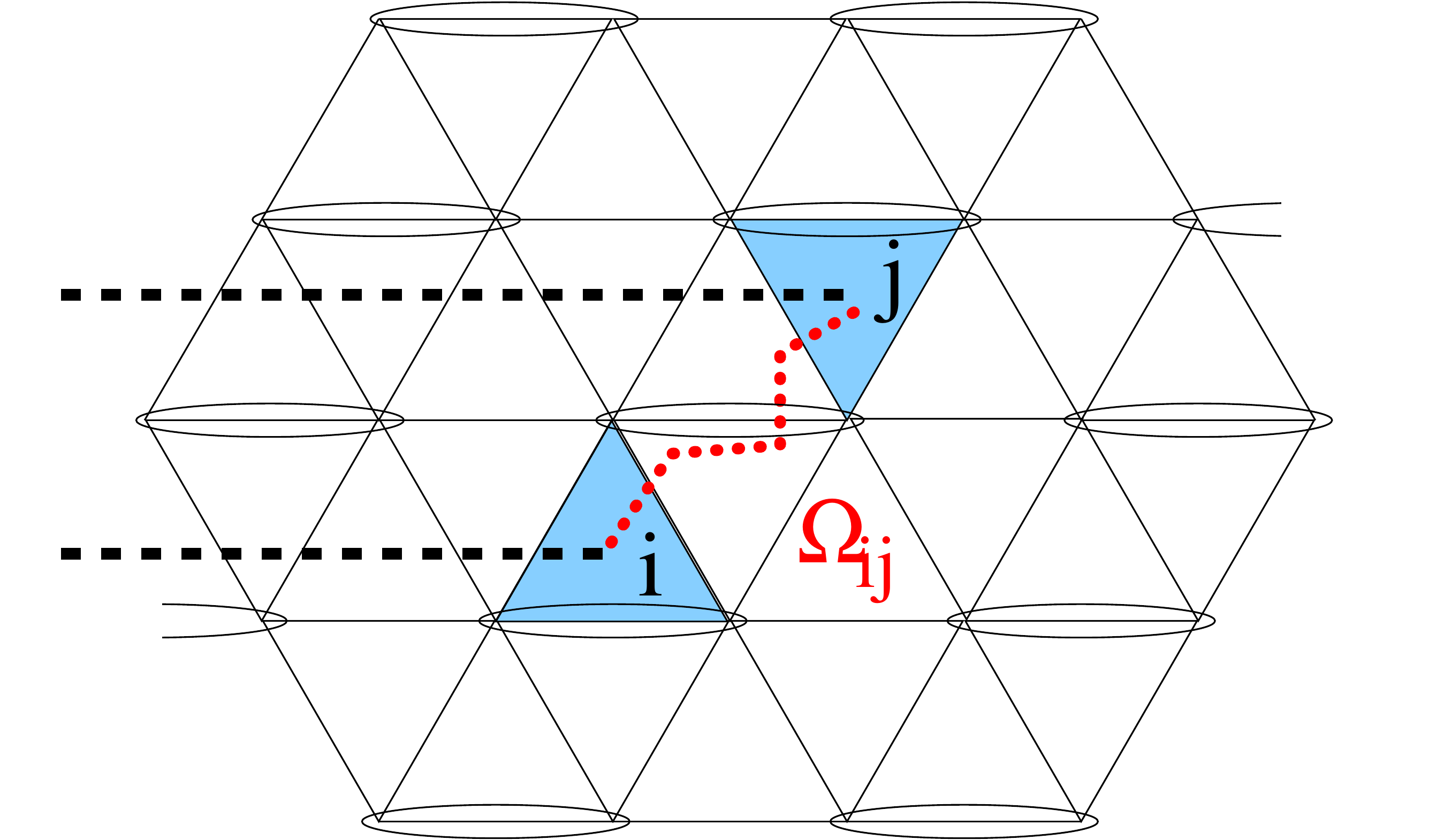}
 \caption{(Color online) Dashed lines: Horizontal paths used in Eq.~\ref{eq:sigmax} connecting the boundary of the system
 to triangles $i$ and $j$.
 Dotted segment: Path $\Omega_{ij}$. In this example $\Omega_{ij}$ crosses one dimer of the reference
 configuration $c_0$ (ellipses) and, therefore, $\epsilon_{ij}=-1$ (Eq.~\ref{eq:epsilon})}
 \label{fig:string}
\end{figure}

This can be achieved by introducing spin-$\frac{1}{2}$ operators $\vec \sigma_i$ on the dual honeycomb lattice:
\begin{equation}
\sigma^x_i=\prod_{l(i)} \tau^z_{l(i)}, \ \ \sigma^z_i =
\prod_{l<i} \tau^x_l,
\label{eq:sigmax}
\end{equation}
where $l<i$ represents all bonds cutting a straight  path (say, horizontal, see Fig.~\ref{fig:string}
or \ref{fig:def}) starting at $i$
(we implicitly assume a finite lattice with open boundary conditions).
Combined with the constraint, this definition implies that
\begin{equation}
 \sigma^z_i \sigma^z_j=M_{ij}\;\tau^x_{l}
 \label{eq:sst}
\end{equation}
for two neighboring
sites $i,j$ separated by the bond $l$. $M_{ij}=\pm1$ is such that each hexagon
has exactly one $M_{ij}=-1$ bond
(see Fig.~\ref{fig:hex}).
In terms of these spin operators, the Hamiltonian is 
the fully frustrated Ising model (FFIM)\cite{villain77} discussed by
Moessner and Sondhi:\cite{ms01}$^,$\footnote{See also Refs.~\onlinecite{js91,sf00,sf01}
for the relation between frustrated Ising models and the $\Z$ gauge theories appearing in the description
of Mott insulators.}
\begin{equation}
H= H_J+H_\Gamma = - J \sum_{\langle i,j \rangle} M_{ij} \sigma^z_i \sigma^z_j - \Gamma \sum_i \sigma^x_i.
\label{eq:imtf}
\end{equation}
Note that choosing other paths to define $\sigma^z_i$ leads to other signs for $M_{ij}$, which,
however, are always such that  an odd number of minus signs appear around each
hexagon, leading to the same Ising model up to a gauge transformation. The present choice leads
to the smallest  unit cell (4 honeycomb sites).

In the following, we study the phase diagram and the excitations of this model
within a semiclassical  (large $S$) approximation.

\subsubsection{Classical phase diagram}
\label{sssec:cpd}

First, we determine the classical ground state of the model as a function
of $\Gamma/J$. To this end, we replace the spin-$\frac{1}{2}$ operators by classical 3-component 
vectors of unit length. Since the $y$ component does not appear in the Hamiltonian, it is 
clear that in the ground state the magnetization must lie in the $x$-$z$ plane. 
To find the lowest energy solution, we use the following parametrization:
\begin{eqnarray}
 \sigma^z_i &\to& \rho(i), \\
 \sigma^x_i &\to& \sqrt{1-\rho(i)^2},
 \label{eq:sigma_rho}
\end{eqnarray}
with $|\rho(i)|\leq 1$.
The corresponding energy is
\begin{equation}
    E=-\frac{J}{2}\sum_{ij} M_{ij} \rho(i) \rho(j) -\Gamma \sum_{i} \sqrt{1-\rho(i)^2}.
    \label{eq:Ecla}
\end{equation}

Initially defined for nearest neighbors (Eq.~\ref{eq:sst}), the coefficients
$M_{ij}$ have been upgraded to a matrix $M$ by setting all other elements to 0.
This matrix describes the motion of a particle on a honeycomb lattice with 4 sites per unit cell
and a flux $\pi$ per hexagonal plaquette.
It reduces to a 4$\times$4 matrix after Fourier transformation.
The eigenvalues associated with the momentum $k=(k_x,k_y)$ are\cite{ms01}
\begin{eqnarray}
  &m_{k}^{1,2,3,4}=&\nonumber\\
  &\pm\sqrt{3\pm\sqrt{2[3+\cos(2k_x)-\cos(k_x+k_y)+\cos(k_x-k_y)]
  }}.&\nonumber
\end{eqnarray}
Likewise, we denote by $\rho$ the column vector of components $\{\rho_i\}$.

{\fbox {$\Gamma/J>\sqrt{6}$}}
For $\Gamma\gg J$, we expect the $\rho(i)$'s to be small (or zero). We can, therefore, expand
Eq.~\ref{eq:Ecla} to quadratic order:
\begin{equation}
    E=-\frac{J}{2} \rho^t M \rho +\frac{1}{2}\Gamma \rho^t \rho
    \label{eq:Ecla2}.
\end{equation}
The largest eigenvalue of $M$ being $\sqrt{6}$, we find
that, as long as $\Gamma/J>\sqrt{6}$, the energy is minimized by $\rho=0$
and all spins point in the $x$ direction.

\begin{figure}
 \includegraphics[height=3.8cm]{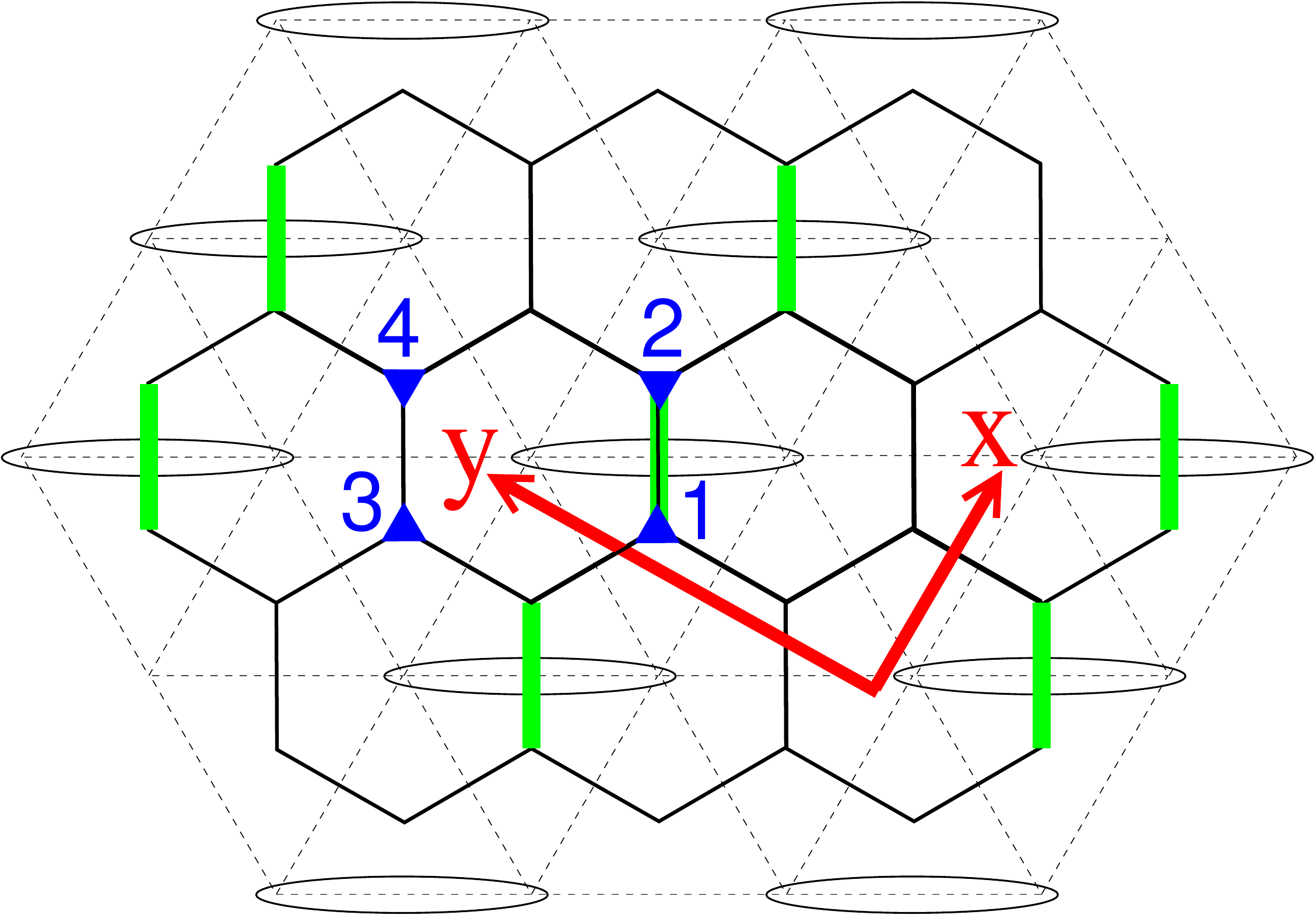}
 \caption{(Color online)
 The fat (green) bonds of the hexagonal lattice (crossed by a dimer of the reference configuration on the
 triangular lattice) have $M_{ij}=-1$, the other bonds
 have $M_{ij}=1$.
 The four sites of the unit cell
 (small blue triangles) are labelled from
 1 to 4 and the Bravais vector are $x$ and $y$.}
 \label{fig:hex}
\end{figure}

\fbox{$0< \Gamma/J\leq \sqrt{6}$}
At $\Gamma/J=\sqrt{6}$, all the real eigenvectors of $M$ for the eigenvalue $\sqrt{6}$
(satisfying $|\rho_i|\leq 1$) minimize Eq.~\ref{eq:Ecla2}.
Let us choose four complex vectors which form a basis of the subspace associated with the eigenvalue $\sqrt{6}$:\cite{ms01}
\begin{eqnarray}
    v_1(x,y)&=&v_3^*=\left[\begin{array}{c}
           e^{5i\pi/12}/F \\
           e^{-i\pi/6}/F \\
           1        \\
           e^{-i\pi/12}
          \end{array}\right] \exp\left(\frac{i\pi}{6}x +
           i\frac{\pi}{2}y\right),\nonumber \\
    v_2(x,y)&=&v_4^*=\left[\begin{array}{c}
           e^{i\pi/12} \\
           e^{-5i\pi/6}\\
           1/F\\
           e^{-5i\pi/12}/F
          \end{array}\right] \exp\left(i\frac{5\pi}{6}x +
    i\frac{\pi}{2}y\right),\nonumber \\
          F&=&2\sin(5\pi/12),
\end{eqnarray}
where the 4 entries of the vector refer to the 4 sites in the unit cell
(numbered as in Fig.~\ref{fig:hex}) and $x,y$ are the (Bravais) coordinates of the unit cell.
These eigenvectors correspond to the four points labelled B in the rectangular Brillouin
zone of Fig.~\ref{fig:bril}.

The most general real eigenvector $\rho$ can be parametrized by three angles $\alpha_1$, $\alpha_2$, and $\beta$
and a normalization factor $\lambda$:
\begin{equation}
    \rho(i)=\lambda\left[
            \cos(\beta) \Re(v_1(i)e^{i\alpha_1}) + \sin(\beta) \Re(v_2(i)e^{i\alpha_2})
        \right].
        \label{eq:rho_angles}
\end{equation}
To find the ground state, the energy has to be minimized as a function 
of these 3 parameters.
The analysis is made easier when one realizes that the second and fourth
moments of the spin deviations are independent
of the three angles. Indeed,
\begin{eqnarray}
    \frac{1}{N}\sum_i \rho(i)^2 &=&r^2\lambda^2, \\
    \frac{1}{N}\sum_i \rho(i)^4 &=&2r^4\lambda^4,\\
    {\rm with}\;\;r^2&=&\frac{1}{4}\left(1+\frac{1}{F^2}\right).
\end{eqnarray}
Replacing $\rho$ by Eq.~\ref{eq:rho_angles} in the expression for the energy
(Eq.~\ref{eq:Ecla}), we get up to order $\lambda^4$:
\begin{equation}
    E/N=-\Gamma+\frac{1}{2}\left(\Gamma-J\sqrt{6}   \right)  r^2\lambda^2+\frac{1}{4}\Gamma r^4\lambda^4
        +\mathcal{O}(\lambda^6)
        \label{eq:Elambda}
\end{equation}
Minimizing with respect to $\lambda^2$ gives:
\begin{equation}
    \lambda^2=\frac{J\sqrt{6}-\Gamma}{f^2 \Gamma}.
    \label{eq:lambda2}
\end{equation}
For $\Gamma> J\sqrt{6}$ the energy is minimized for $\lambda=0$. For $\Gamma\leq J\sqrt{6}$,
we have to expand Eq.~\ref{eq:Ecla} to the 6$^{\rm th}$ order to find the angles
$\alpha_1$, $\alpha_2$ and $\beta$ which minimize the classical energy. Indeed,
Eq.~\ref{eq:Elambda} shows that, {\it up to order $\lambda^4$,
the energy is independent of the three angles}. At this order, the energy is constant and minimum on a 3-dimensional sphere,
a consequence of the $O(4)$ symmetry discovered
by Moessner and Sondhi.\cite{ms01} One, therefore, has to go to the next order to find the actual minima. 
At 6$^{\rm th}$ order in $\lambda$, the energy is minimized when
\begin{equation}
\frac{1}{N}\sum_i \rho(i)^6
\end{equation}
is minimum. A numerical
investigation shows that the solutions
(for $\Gamma/J$ close to but below $\sqrt{6}$) are 48-fold degenerate
and can be deduced from each other by symmetry operations
(12 translations and 4 point-group operations).
This confirms the result obtained previously on symmetry
grounds.\cite{msc00,ms01}

\begin{figure}
\includegraphics[width=6.8cm]{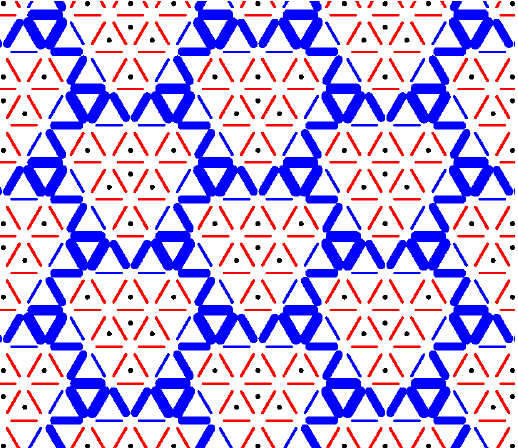}\vspace*{0.3cm}
 \caption{
 (Color online) Plot of the rescaled relative ``dimer'' density $D_{ij}$ in one of the 48
 classical ground states of the FFIM in transverse field,
 for a transverse field $\Gamma$ just below $\Gamma_c$.
 Thin red bonds represent $D_{ij}=0$ (corresponding to the {\it highest} ``dimer density'' $d_{ij}=0.5$).
 Blue bonds are for $D_{ij}<0$, that is, a {\it lower} dimer density, with a width proportional to $|D_{ij}|$.
 $D_{ij}$ takes only 4 different values: 0, -0.259, -0.518, and -0.776.
 }
 \label{fig:vbc} 
\end{figure}

Motivated by the FFIM/dimer model correspondence, we are interested in the average
``dimer density'':
\begin{equation}
 d_{ij}=\frac{1}{2}(1-M_{ij} \langle\sigma^z_i \sigma^z_j\rangle)
\end{equation}
for all pairs $(i,j)$ of nearest-neighbor honeycomb sites.
In the classical limit, this may be approximated by
\begin{equation}
 d_{ij}=\frac{1}{2}(1-M_{ij}\rho(i)\rho(j)),
\end{equation}
where $\rho$ (Eq.~\ref{eq:rho_angles}) is the classical ground state.
Close to the  transition at $\Gamma_c=J\sqrt{6}$,
$\rho$ scales as $\lambda\sim \sqrt{\Gamma_c-\Gamma}$
(Eq.~\ref{eq:lambda2})
and the ``dimer density'', thus, shows small deviations about $\frac{1}{2}$.
To visualize the ``dimerization'' pattern in the vicinity of $\Gamma_c$,
the appropriate
quantity is, therefore, a relative rescaled ``dimer'' density defined by
$D_{ij}=\frac{1}{\lambda^2}(d_{ij}-\frac{1}{2})$, and
plotted in Fig.~\ref{fig:vbc}
for one of the 48 ground states. The obtained pattern is 
highly reminiscent of the $\sqrt{12}\times\sqrt{12}$ VBC observed
in the triangular lattice QDM.
In particular, the 48 spin configurations give only 12 different ``dimer'' patterns, with a smaller
unit cell containing 12 sites of the triangular lattice.
It is also interesting to notice that $\rho(i)$ vanishes
in the triangles located inside the large diamonds (marked with a dot in Fig.~\ref{fig:vbc}).
From Eq.~\ref{eq:sigma_rho}, the corresponding sites have a magnetization pointing exactly in the $x$ direction,
which means a total absence of ``vison'' and local correlations identical to that of the $J=0$ paramagnetic
ground state.
The ``dimer'' density $d_{ij}$ is {\it uniform} inside
these diamonds, in qualitative agreement with the intriguing observation made in Ref.~\onlinecite{ralko06}.

\subsubsection{Semi-classical analysis for $\Gamma/J\ge \sqrt{6}$}
\label{sssec:sca}

If $J=0$, the ground state is the fully polarized state $\sigma^x_i=1$. The
first (degenerate) excited state is obtained by flipping one spin at some arbitrary triangle. To first order in $J/\Gamma$,
the spectrum is obtained by diagonalizing the perturbation $H_J$ in this subspace.
This amounts to diagonalizing $M$, and leads to the eigenvalues 
$\epsilon^i_k=2\Gamma + Jm^i_k$, a result already obtained previously.\cite{ms01}

\begin{figure}
 \hspace*{-1.5cm}\includegraphics[width=7cm]{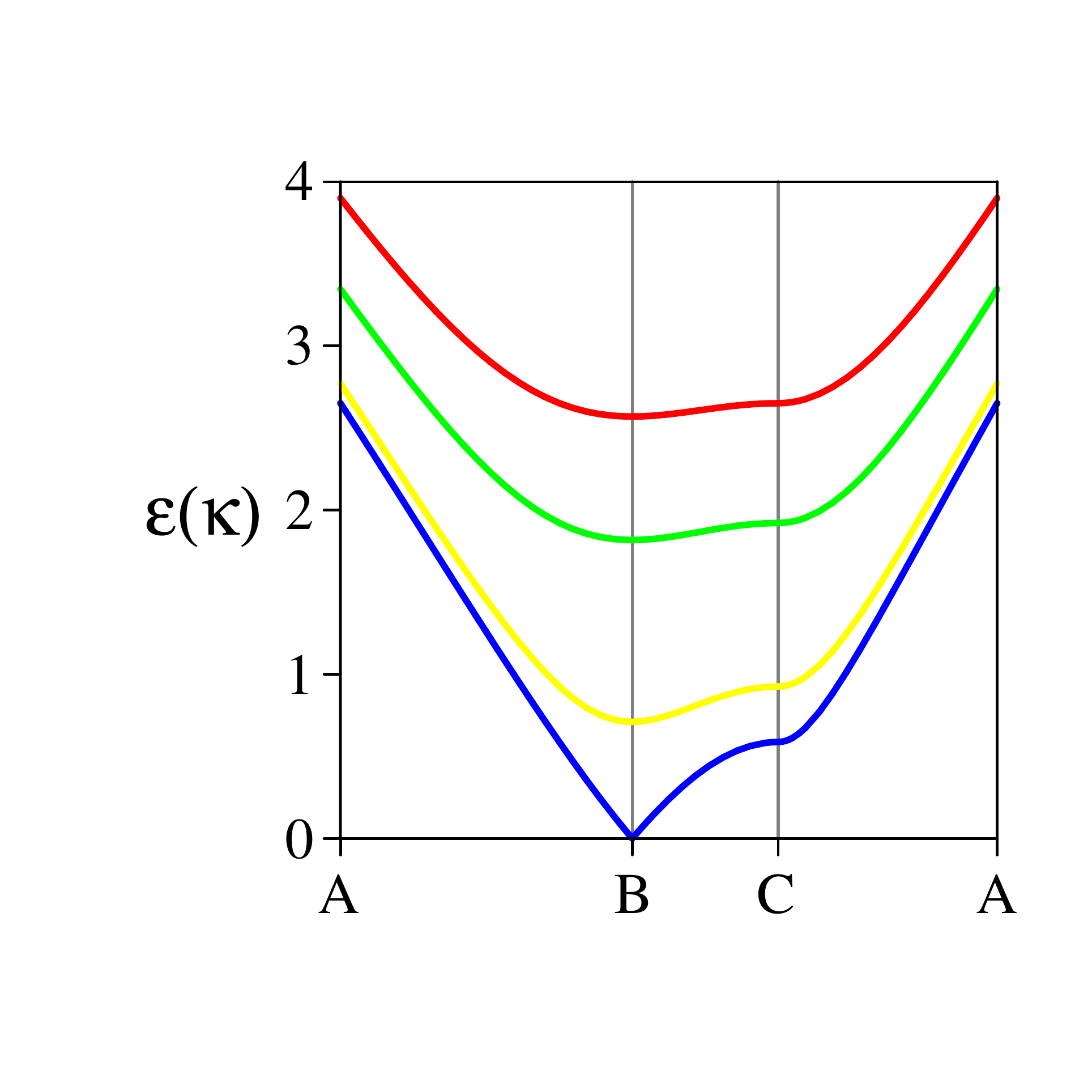}\vspace*{-0.7cm}
 \caption{(Color online) Spin-wave dispersion relation $S\epsilon_k/J$ (Eq.~\ref{eq:spinwave})
 for 
 the path A$\to$B$\to$C$\to$A in the Brillouin zone (see Fig.~\ref{fig:bril})
 and different values of $\Gamma/J$:3 (top), 2.75, 2.5, and $\sqrt{6}\simeq2.448$
 (bottom). At the critical point $\Gamma/J=\sqrt{6}$,
 the spectrum is linear around $k=(\pi/6,\pi/2)$ and $k=(5\pi/6,\pi/2)$.} 
 \label{fig:sw}
\end{figure}

\begin{figure}
 \hspace*{0.8cm}\includegraphics[width=5.3cm]{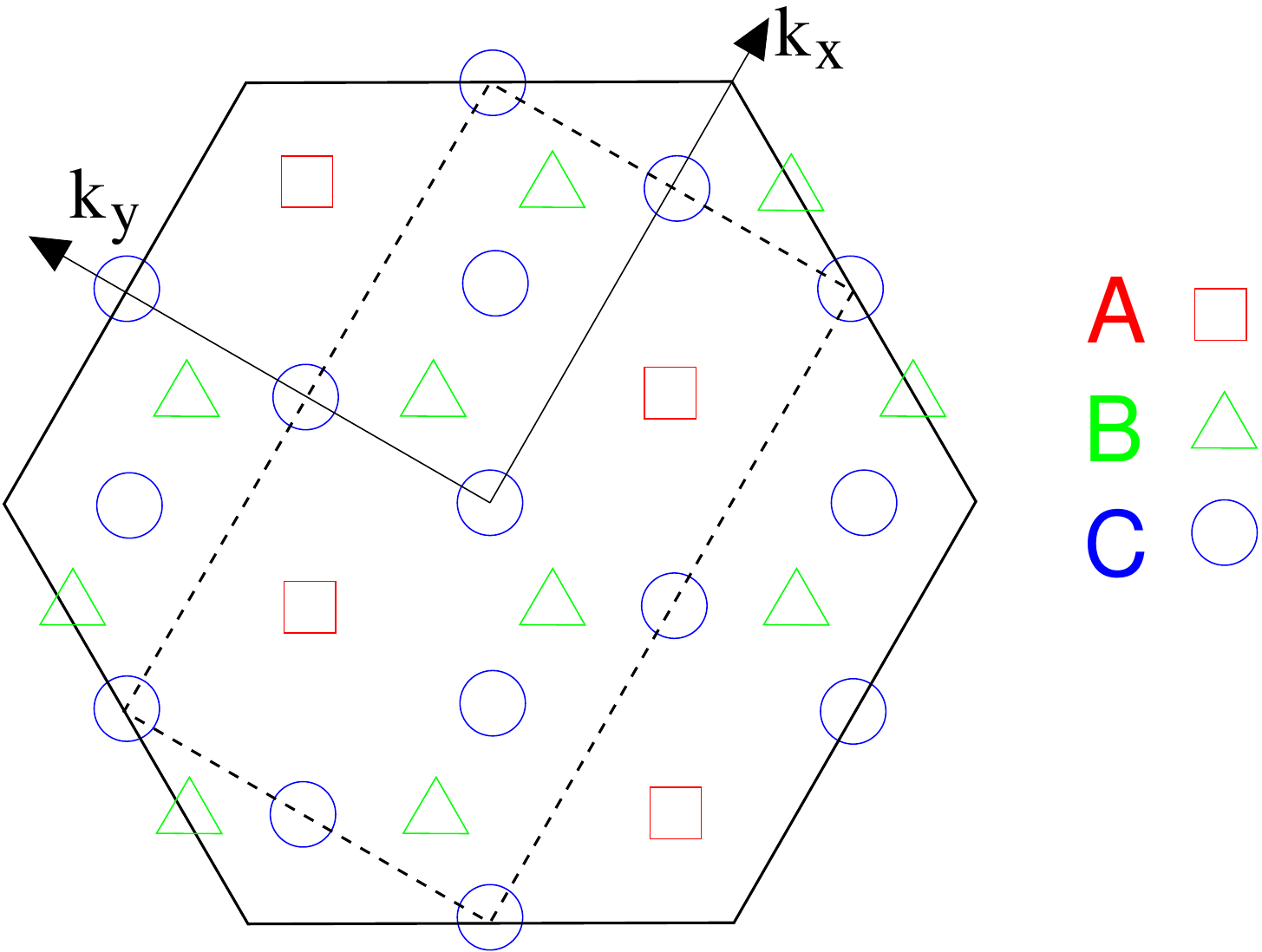}
 \caption{
 (Color online) Dashed rectangle: Brillouin zone ($-\pi \leq k_x \leq \pi$ and $-\pi \leq k_y \leq \pi$)
 of the hexagonal lattice shown in Fig.~\ref{fig:hex}. A, B, and C are the high-symmetry points used in
 Fig.~\ref{fig:sw}. The large hexagon is the Brillouin zone of the underlying triangular lattice.
 }
 \label{fig:bril}
\end{figure}

To go further, beyond the  limit $J/\Gamma\ll1$, we perform
a $1/S$ semiclassical expansion. 
We generalize the spin-1/2 Hamiltonian to an arbitrary value $S$ of the spin:
\begin{equation}
    \mathcal{H}=-\frac{J}{S^2}\sum_{\langle i,j\rangle} M_{ij} S^z_i S^z_j -\frac{\Gamma}{S}\sum_i S^x_i
\end{equation}
(which reduces to Eq.~\ref{eq:imtf} when $S=1/2$).
Spin deviations away from the $x$ directions are represented using Holstein-Primakoff bosons.
To leading order in $1/S$:
\begin{eqnarray}
S^z_i&=&\frac{1}{2}\sqrt{2S}\left(b^\dagger_i+b_i\right), \\
S^x_i&=&S-b^\dagger_i b_i.
\end{eqnarray}
One truncates the Hamiltonian to quadratic order in $b_i$, and diagonalizes it
through a Bogoliubov transformation:
\begin{eqnarray}
    b^\dagger_i&=&\sum_j U_{ij} a^\dagger_j +\sum_j V_{ij} a_j
\end{eqnarray}
For the $ a^\dagger_j$ operators to be bosonic creation operators, the matrices $U$ and $V$
must satisfy $U^\dagger U-V^\dagger V=1$.
It is convenient to introduce a unitary matrix $\Omega$ which transforms
$M$ into a diagonal matrix $\tilde M$:
\begin{eqnarray}
    M& =&\Omega \tilde M \Omega^\dagger, \\
       \tilde M_{kp} &=& m_k \delta_{kp}.
\end{eqnarray}
In this new basis, we can look for diagonal solutions for $U$ and $V$:
\begin{eqnarray}
    U=\Omega \tilde U \Omega^\dagger \;\;,\;\;
    V=\Omega \tilde V \Omega^\dagger, \\
       \tilde U_{kp} = u_k \delta_{kp} \;\;,\;\;
     \tilde V_{kp} = v_k \delta_{kp}.
\end{eqnarray}
Since $M$ is real, we can choose $\Omega\in O(N)$.
After some algebra, one finds that the $u_k$ and $v_k$ which diagonalize $H$
are given by
\begin{eqnarray}
    u_k&=&\cosh(\theta_k), \\
    v_k&=&\sinh(\theta_k), \\
    \tanh(2\theta_k)&=&-\frac{m_k}{m_k+2\Gamma/J}.
\end{eqnarray}
The energies of the Bogoliubov excitations are given by
\begin{eqnarray}
    \epsilon_k&=&\frac{J}{S}\sqrt{\frac{\Gamma}{J}\left(m_k+\frac{\Gamma}{J}\right)}.
    \label{eq:spinwave}
\end{eqnarray}
As expected, the spectrum becomes gapless at $\Gamma/J=\sqrt{6}$
and we recover the localized spin flip energy $\epsilon_k\simeq \Gamma/S$ when $\Gamma\to\infty$.
The Bogoliubov spectrum is shown Fig.~\ref{fig:sw} for a few values of $\Gamma$ at or above
$\Gamma_c$. One clearly sees that the dispersion is linear around the
gapless points at the transition.

This behavior is remarkably similar to that found numerically by Ralko {\it et al.}\cite{ralko07} for the QDM by Green's function 
Quantum Monte Carlo. In the next section, we build on this resemblance to develop a variational
approach to the vison spectrum of the QDM.
Note that, as stated above, the $\Z$ gauge theory and the QDM can only be rigorously mapped
onto each other deep into the confined (resp. VBC) phase, and the resemblance between their
spectra in the deconfined (resp. RVB) phases might seem fortuitous. A somewhat deeper connection
will be described in the next section.

\section{Visons in the Quantum Dimer Model}

In the Ising model, elementary excitations for large enough $\Gamma/J$ are spin 
flips, induced by $\sigma^z_i$, which delocalize and get dressed under the effect of $H_J$. 
In the equivalent dual gauge theory, this excitation is produced by the nonlocal string operator 
$\sigma^z_i=\prod_{l<i} \tau^x_l$.
By analogy, it is natural to define a ``point'' vison creation operator for the QDM  by\cite{rc89,ioselevich02}
\begin{equation}
V_{i} = (-1)^{\hat N(i)}
\;\;,\;\;\hat N(i)=\sum_{l<i} \hat n_l,
\label{eq:bare_vison}
\end{equation}
where the dimer operator $\hat n_l$ is defined by $\hat n_l=1$ if bond $l$ is occupied, and $0$ otherwise.
Again, we consider here a finite lattice with open boundary conditions.
Let us first see to which extent this operator is the analog of the vortex creation operator in Ising
gauge theories.

\subsection{$\Z$ gauge structure of QDM}

In gauge theories, the Wilson loop operator defined by 
\begin{equation}
W_{\partial\Omega}=\prod_{l\in\partial\Omega}\tau^z_l
\end{equation}
plays a central role since it allows one to distinguish (in the absence of matter field, as here) the deconfined and confined phases depending on
whether its ground state expectation value (or flux) tends to zero exponentially
with the perimeter of the domain $\Omega$ (deconfined) or with
the area of the domain (confined).
For a $\Z$ gauge theory $W_{\partial\Omega}^2=1$
and the flux going through $\Omega$ measured by $W_{\partial\Omega}$ can only take two values $\pm 1$.
In that respect, an interesting property of the 
operator $\sigma^z_i=\prod_{l<i} \tau^x_l$ is that it changes the flux
between -1 and +1 
if the site $i$ is inside the domain.
Then, $\sigma^z_i \sigma^z_j$ commutes with the Wilson loop and
does not change 
the flux unless $i$ and $j$ sit on {\it opposite sides} of the boundary.
Since the deconfined phase can be accessed from the $J\ll \Gamma$ limit,
the ground state can be thought of as an Ising paramagnet
($\sigma^x=1$ and $W_{\partial\Omega}=1$ everywhere) ``dressed'' perturbatively 
by successive applications of $J\sigma^z_i \sigma^z_j$, where $i$ and $j$
are nearest neighbors. Since $J\sigma^z_i \sigma^z_j$ only changes the flux 
for pairs $ij$ {\it across the boundary}, the expectation value of the
Wilson loop operator behaves according to a {\it perimeter} law.
A similar perturbative argument can be used to derive the area law in the confined phase.

Most of these standard $\Z$ gauge theory results apply to the QDM, with one difficulty however. The Wilson loop operator
cannot be defined in the same way since flipping the dimer occupation along a loop
will often lead to an unphysical state that violates the condition of having
exactly one dimer emanating from each site. Different ways to overcome this problem
can be envisaged. One possibility is to accept that the Wilson loop operator gives zero
when applied to states that are non-flippable along the loop. The property
$W_{\partial\Omega}^2=1$ is lost, but 
$W_{\partial\Omega}$ has eigenvalues $0$, $+1$, or $-1$, and 
eigenstates with eigenvalues $+1$ or $-1$ are still interchanged when a vison operator
(Eq.~\ref{eq:bare_vison})
is applied inside the domain. So confinement or deconfinement of visons is still
expected to lead to area or perimeter laws.

Alternatively, starting from the observation that, when it does not give zero, 
the Wilson loop operator just shifts the dimers along the contour, one can try to 
define a (complicated) flux operator in the dimer space which
shifts the dimer along a ``fattened'' contour which ``adapts'' locally to the dimer configuration 
it acts upon. Although tricky to define in practice,\footnote{On the kagome lattice (and on any lattice made of corner sharing triangles in general), this can be done explicitly in a simple way.\cite{msp02}} this operator would have the advantage that
its square is still equal to 1, preserving the manifest $\Z$ structure of the theory.

Another way to underline the deep connection between the two models is to
consider the Ising model as a ``soft-dimer'' model and
introduce the projection
operator onto the
``hard-core'' dimer
Hilbert space, defined by:
\begin{equation}
    \hat P = \prod_p \frac{(\hat n_p-3)(\hat n_p-5)}{(1-3)(1-5)}
\end{equation}
where the operator $\hat n_p$ counts the number of frustrated bonds (dimers) around the plaquette $p$:
\begin{equation}
    \hat n_p=\frac{1}{2}\sum_{i=1}^6  (1-M_{i,i+1} \sigma^z_i \sigma^z_{i+1})
\end{equation}
By construction, $\hat P=1$ on all the Ising configurations which correspond to a valid
``hard-core"
dimer covering (of the triangular lattice) and  $\hat P=0$ otherwise.
Now, $\hat P$ commutes with $\prod_i \sigma^x_i$ since all terms
in $\hat P$ contain an even number of $\sigma^z$ operators.
This means that $\hat P$ conserves the total flux, so that a spin state
with -1  flux  becomes a dimer wave function with an odd number of visons after
projection.

\subsection{Point vison}

Equation~\ref{eq:bare_vison} defines the simplest operator which changes
the flux, that is, which creates a vison.  So we may consider
\begin{eqnarray}
  |i\rangle &=& V(i) | RK\rangle =\frac{1}{\sqrt{\mathcal N}} \sum_c (-1)^{\hat N(i)} | c \rangle \label{eq:point_vison} \\
    &=&\frac{1}{\sqrt{\mathcal N}} \hat P\sigma^z_i | \left\{ \sigma^x_j=+1 \right\} \rangle \label{eq:Psigmaz}
\end{eqnarray}
as a first variational approximation
to the true lowest eigenstate of $\mathcal{H}$ in the $-1$ flux sector
($\mathcal N$ is the total number of dimer coverings).
The ground state $|RK\rangle$ is a zero-energy eigenstate of $\mathcal{H}$,
implying that
the expectation value $E_K<0$ of the kinetic energy term exactly compensates
the expectation value
$E_P>0$ of the potential energy term.
Indeed,
\begin{equation}
    E_P= 3 N p_0 = -E_K,
\end{equation}
where $3N$ is the total number of rhombi ($N$ the number of sites) and $p_0$ the probability
to have 2 parallel dimers on a given rhombus in the classical dimer problem (with uniform measure over all dimer
configurations). Using the Pfaffians method,\cite{k61,krauth} one finds in the thermodynamic limit 
\begin{equation}
    p_0\simeq 0.0933310104...
\end{equation}

In $|i\rangle$ the expectation value of the potential energy term is the same as in $|RK\rangle$.
In fact, all the observables which are diagonal in the dimer basis commute with $\sigma^z$
and thus have the same expectation value in the ground state $|RK\rangle$ and in the trial wave function $|i\rangle$.
The increase of the energy is only kinetic. Because the kinetic terms corresponding to the three diamonds around $i$
anticommute with $\sigma^z_i$,  their expectation values has changed sign.
Thus we find
\begin{eqnarray}
    \langle i | \mathcal{H} | i \rangle
    -	\langle RK | \mathcal{H} | RK \rangle
    = 6 p_0 \simeq  0.56.
\end{eqnarray}

\subsection{Dispersing point vison}
\label{sec:vison0}

One can lower the energy by constructing plane waves. To compute
the associated dispersion relation we have to evaluate the following matrix elements:
\begin{eqnarray}
 S_{ij}^0&=&\langle i | j \rangle       \label{eq:Sdef} \\
 H_{ij}^0&=&\langle i | \mathcal{H} | j \rangle \label{eq:Hdef} 
\end{eqnarray}
We begin with the overlap matrix
\begin{eqnarray}
 S_{ij}^0&=&\frac{1}{\mathcal N} \sum_c \langle c | (-1)^{\hat N(i)+\hat N(j)} | c \rangle
\end{eqnarray}
($\hat N(i)$ and $\hat N(j)$ are defined in Eq.~\ref{eq:bare_vison})
which simplifies to
\begin{eqnarray}
 S_{ij}^0&=&\epsilon_{ij} \frac{1}{\mathcal N} \sum_c \langle c |(-1)^{\hat N(i,j)}| c \rangle    \label{eq:S} \\
 \epsilon_{ij}&=&(-1)^{N_0(i,j)}
    \label{eq:epsilon}
\end{eqnarray}
where the {\it local} operator $\hat N(i,j)$ counts the number of dimers across
some path $\Omega_{ij}$ connecting the triangles $i$ and $j$
(see Fig.~\ref{fig:string}), and $N_0(i,j)$ is equal to
that number in the reference configuration $c_0$,
chosen with all the dimers horizontal (Fig.~\ref{fig:string}).
This follows from two simple properties: $\langle c | (-1)^{\hat N(i)+\hat
  N(j)+\hat N(i,j)} | c \rangle$
is independent of the configuration $c \rangle$, and $\langle c_0 | (-1)^{\hat N(i)+\hat
  N(j)} | c_0 \rangle=1$.
We finally write
\begin{equation}
    S_{ij}^0=\epsilon_{ij} \left< (-1)^{\hat N(i,j)} \right>
    \label{eq:Saverage}
\end{equation}
where $\left< \cdots \right>$ represents the average with equal weight over all dimer coverings.

The average of any such diagonal observable can be computed using
the Pfaffian of a (modified) Kasteleyn matrix.\cite{k61,krauth}
In the present case, the ``string'' observable $(-1)^{\hat N(i,j)}$
is coded in the Kasteleyn matrix by changing the signs of the matrix elements corresponding to
bonds crossed by $\Omega_{ij}$, {\it i.e.}, setting some bond fugacities to $-1$.
To evaluate numerically such an expectation value, we construct the modified Kasteleyn matrices
corresponding to a large enough triangular lattice, in which the path $\Omega_{ij}$
is embedded. Finite-size effects decay exponentially with the system size, so that
lattices with $28\times28$ sites (with periodic boundary conditions) can safely be used to evaluate
$S_{ij}^0$ with high accuracy up to distances $d\simeq 10$ between triangles $i$ and $j$.

\begin{figure}
 \includegraphics[width=7cm]{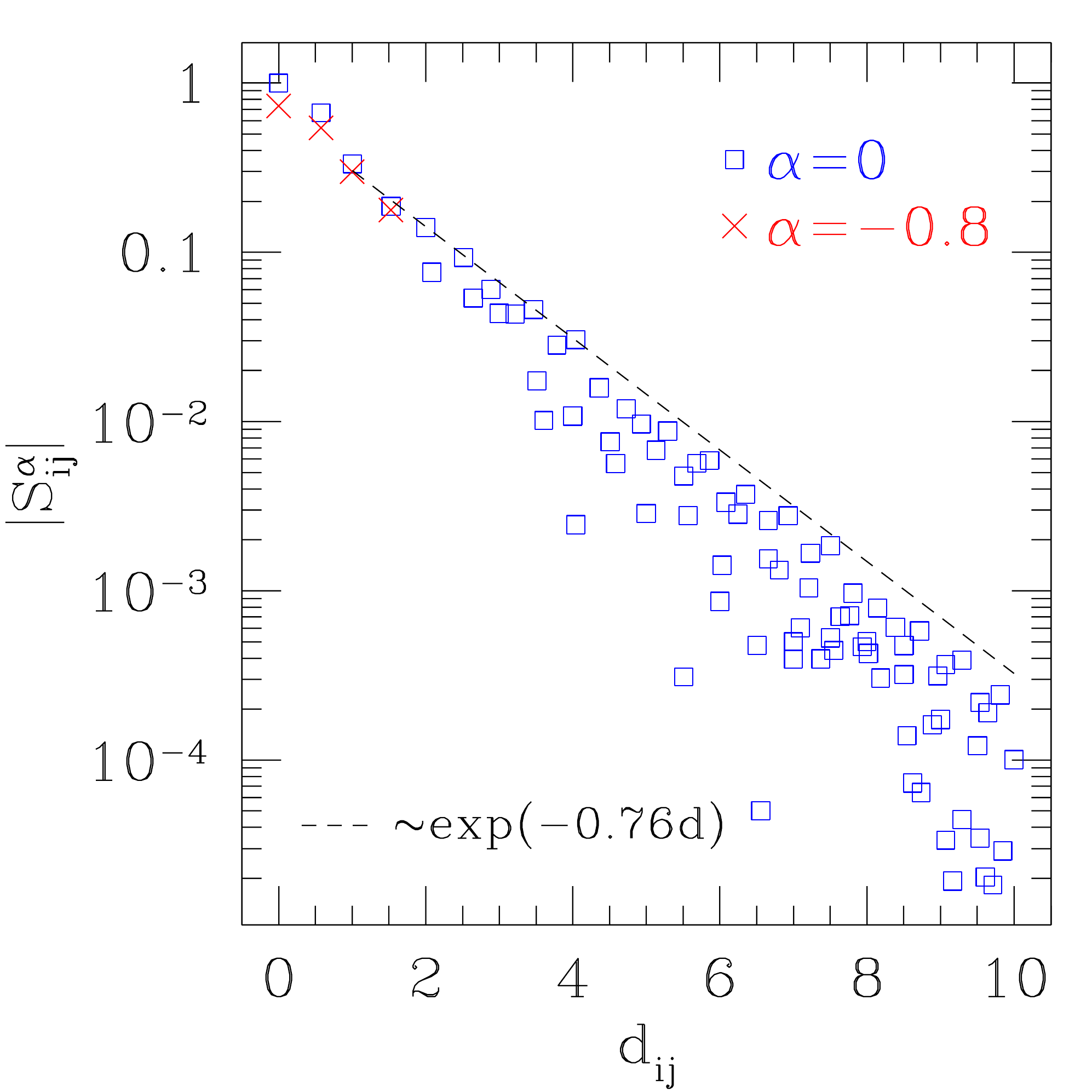}
 \caption{(Color online) Overlap $|\langle i|j\rangle|$ between two point-vison states
 (defined in Eq.~\ref{eq:point_vison}) as a function the distance $d_{ij}$ between $i$ and $j$.
 The (blue) square correspond to Eq.~\ref{eq:Sdef}
 and the three (red) crosses correspond to
 Eq.~\ref{eq:Sija} with $\alpha=-0.8$. The calculations are done using
 an exact evaluation of the Pfaffians on a $28\time 28$-site lattice with periodic boundary conditions.
 Dashed line: guide to the eye corresponding to an exponential decay with
 the dimer-dimer correlation length $\xi^{-1}=0.76$.\cite{ioselevich02}
 }
 \label{fig:sij}
\end{figure}

The matrix elements $S_{ij}^0$, plotted in Fig.~\ref{fig:sij} as a function of the distance
between $i$ and $j$, decay exponentially (a result anticipated
by Read and Chakraborty\cite{rc89}).
These quantities have already been evaluated by Ioselevich {\it et al.}\cite{ioselevich02} using a classical
Monte Carlo sampling.

\begin{figure}
 \includegraphics[width=11cm]{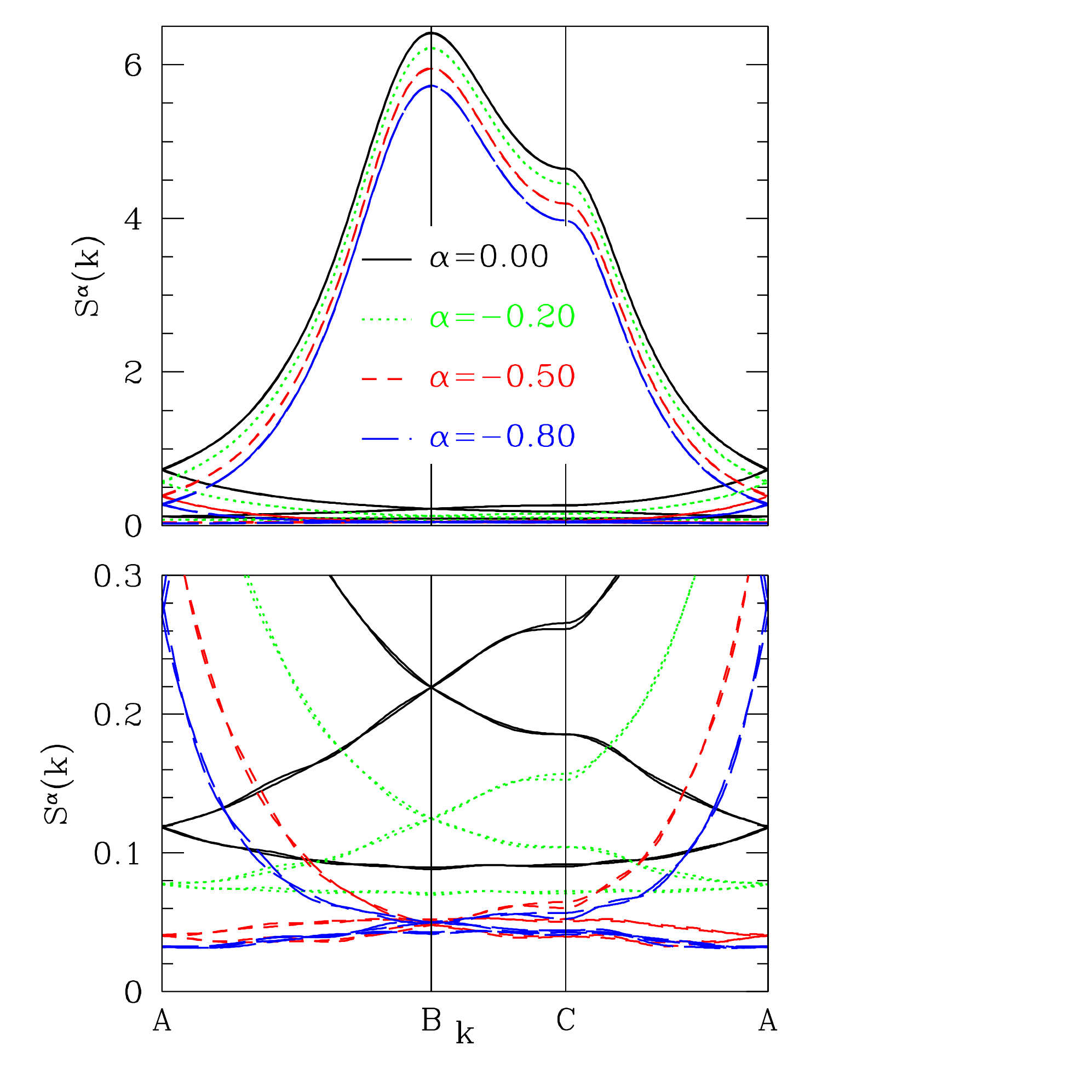}\hspace*{-2cm}
 \caption{
  (Color online)
  Spectrum of the overlap matrix $S^\alpha(k)$ (Eqs.\ref{eq:Sdef} and \ref{eq:Sija}), for different values
of $\alpha$ (0, -0.2, -0.5 and -0.8),
along the path  A$\to$B$\to$C$\to$A in the Brillouin zone (see Fig.~\ref{fig:bril}).
The bottom panel is a zoom on the lowest eigenvalues of $S^\alpha$.
In these calculations, the matrix elements $S_{ij}^\alpha$ are neglected for triangles $i$ and $j$
at distance $d\geq d_{\rm max}=10$
(102$^{\rm th}$ neighbor on the hexagonal lattice).
Up to this distance, the finite-size lattice used in the calculation ($28\times28$ sites)
gives practically the infinite volume limit for $S_{ij}^\alpha$.
The value of $d_{\rm max}$ used here
is large enough to ensure
a good convergence of the spectrum since the curves obtained with a smaller truncation distance ($d_{\rm max}\simeq 8.47$ -- shown here with the same colors)
are almost superposed with that for $d_{\rm max}=10$,
except for the bottom of the spectrum at $\alpha=-0.5$ and $\alpha=-0.8$.
The overlap spectrum turns out to be gapped for $\alpha=0$ and $\alpha=-0.2$
(and probably at $\alpha=-0.5$ and $\alpha=-0.8$ too), indicating
the linear independence of the vison states.}
 \label{fig:sk}
\end{figure}

By construction, a product like $\epsilon_{i_1,i_2} \epsilon_{i_2,i_3} \cdots \epsilon_{i_n,i_1}$
(where $i_1,i_2,\cdots,i_n$ form a closed loop of triangles)
is equal to the parity of the number of sites enclosed in the loop.
Thus, the signs of the matrix elements $S_{ij}^0$ are similar to those of the hopping amplitude
of a particle moving on the hexagonal lattice and subjected to a magnetic
field corresponding
to half a flux quantum per hexagon.\footnote{The fact that the Hamiltonian
describing the vison hopping is not translation-invariant is due to the fact that the
dimer liquid has a non-trivial {\it projective symmetry group} (PSG).\cite{wen02}
}
In such a case, the magnetic unit cell has to be doubled compared to the original lattice cell.
Since the original unit cell contains one triangular site and two triangular plaquettes, 
the magnetic one contains four
triangular plaquettes and thus four sites of the hexagonal lattice.
In Fourier space, $S^0(k)$ is  a $4\times4$ matrix, as the matrix $M$ discussed previously. The same is also true for the Hamiltonian matrix elements described below.

The spectrum of the overlap matrix $S^0(k)$ is plotted in Fig.~\ref{fig:sk}
(for $k$ describing a representative path in the Brillouin zone).
As an important result, the eigenvalues are {\it strictly positive} for all $k$.
To our knowledge, it is the first time
that the {\it linear independence} of point vison states is proved.


Let us now turn to the matrix elements of the Hamiltonian.
We wish to transform Eq.~\ref{eq:Hdef} into an expression which can be evaluated using the Pfaffians, that
is the expectation value of a diagonal observable in the dimer basis.
First, we write $\mathcal{H}$ as a sum of projectors
\begin{eqnarray}
    \mathcal{H}         &=& 2 \sum_{r_0} \hat\Pi_{r_0} \\
    \hat\Pi_{r_0}           &=& |\psi_{r_0} \rangle \langle \psi_{r_0} | \\
    | \psi_{r_0} \rangle    &=& \frac{1}{\sqrt{2}}\left( \left|\rhombH\right> - \left|\rhombV\right> \right )
\end{eqnarray}
and expand Eq.~\ref{eq:Hdef} into
\begin{eqnarray}
    H_{ij}^0&=&\frac{2}{\mathcal N} \sum_{c_1,c_2} \sum_{r_0} 
        (-1)^{N(c_1,i)+N(c_2,j)} \langle c_1 | \hat\Pi_{r_0} | c_2 \rangle,
\end{eqnarray}
with the notation
$
N(c,i)=\langle c |\hat N(i)|c\rangle=\pm1
$.
Because of the double sum $\sum_{c_1,c_2}$, this does not yet have the form
of a diagonal observable amenable to an evaluation with Pfaffians.
To go further we note that 
$ \langle c_1 | \hat\Pi_{r_0} | c_2 \rangle$ vanishes if $c_1$ and/or $c_2$ is not flippable
around the rhombus ${r_0}$. In addition, $\langle c_1 | \hat\Pi_{r_0} | c_2 \rangle=0$ if $c_1$
and $c_2$ differ anywhere outside ${r_0}$. So we can restrict the sum to
pairs  of configurations $(c,\bar c)$ which differ by a dimer flip in ${r_0}$
and which are identical elsewhere on the lattice:
\begin{eqnarray}
    H_{ij}^0&=&\frac{2}{\mathcal N} \sum_{r_0} \sum_{\begin{array}{c}(c,\bar c)\\F_{r_0}(c)=1\end{array}} \nonumber \\
        &&\left[ (-1)^{N(c,i)}\langle c| + (-1)^{N(\bar c,i)}\langle \bar c| \right] \nonumber\\
        &&\hat\Pi_{r_0} 
           \left[ (-1)^{N(c,j)}| c \rangle  + (-1)^{N(\bar c,j)} | \bar c \rangle \right]
\end{eqnarray}
where $F_{r_0}(c)=1$ if $c$ is flippable on rhombus ${r_0}$, and $F_{r_0}(c)=0$ otherwise.

$c$ and $\bar c$ only differ inside ${r_0}$, so the signs
$(-1)^{N(c,i)}$ and $(-1)^{N(\bar c,i)}$ are the same if
$i$ is not inside $r_0$, and are opposite if $i\in {r_0}$.
Let us note $\delta_{i,{r_0}}=-1$ if $i\in {r_0}$ and  $\delta_{i,{r_0}}=1$ otherwise.
This leads to
\begin{eqnarray}
    H_{ij}^0&=&\frac{2}{\mathcal N}
    	\sum_{r_0}
    	\sum_{\begin{array}{c}(c,\bar c)\\F_{r_0}(c)=1\end{array}} 
        (-1)^{N(c,i)+ N(c,j)} \nonumber \\
        &&\times\left[ \langle c| + \delta_{i,{r_0}}\langle \bar c| \right]
        \hat\Pi_{r_0} 
        \left[ | c \rangle  + \delta_{j,{r_0}} | \bar c \rangle \right].
\label{eq:Hij0}
\end{eqnarray}
Using the explicit form of $\hat\Pi_{r_0}$, we get
\begin{eqnarray}
        \left[ \langle c| + \delta_{i,{r_0}}\langle \bar c| \right]
        \hat\Pi_{r_0} 
        \left[ | c \rangle  + \delta_{j,{r_0}} | \bar c \rangle \right]
        =\frac{1}{2}(1-\delta_{i,{r_0}}) (1-\delta_{j,{r_0}})
        \nonumber\\
\end{eqnarray}
and, finally,
\begin{eqnarray}
    H_{ij}^0&=&
    \sum_{r_0}(1-\delta_{i,{r_0}}) (1-\delta_{j,{r_0}})
    \nonumber\\
    &&\times\frac{1}{
    2
    \mathcal N}\sum_{\begin{array}{c}c\\F_{r_0}(c)=1\end{array}} 
	(-1)^{N(c,i)+ N(c,j)}
         \label{eq:Hij0b}\\
        &=&\epsilon_{ij}
        \sum_{r_0} (1-\delta_{i,r_0}) (1-\delta_{j,{r_0}})
        \nonumber\\
        &&\times\frac{1}{2\mathcal N}
        \sum_{\begin{array}{c}c\\
       F_{r_0}(c)=1\end{array}}  \langle c | (-1)^{\hat N(i,j)} | c\rangle.
\end{eqnarray}
The last expression is the average of a diagonal observable and can, thus, be  evaluated using Pfaffians:
\begin{eqnarray}
    H_{ij}^0&=&\frac{1}{2}\epsilon_{ij} \sum_{r_0} (1-\delta_{i,{r_0}}) (1-\delta_{j,{r_0}})
    \nonumber\\&&\times
    \left< 
            (-1)^{\hat N(i,j)} \hat F_{r_0}
        \right>
\end{eqnarray}
where we used an operator notation for the ``flippability''
$\hat F_{r_0}|c\rangle=F_{r_0}(c)|c\rangle$.
To get a nonzero contribution, there must be at least one rhombus ${r_0}$ containing
$i$ and $j$. So $H_{ij}^0=0$ if $i$ and $j$ are not nearest neighbors.
Using modified Kasteleyn matrices in a way similar to that described for the overlap matrix $S^0$,
the nonzero matrix elements can be calculated. We find
$|H_{ij}^0|=6p_0$ for $i=j$,
$|H_{ij}^0|=2p_0$  when $i$ and $j$ are first neighbors.

\begin{figure}
 \includegraphics[width=7cm]{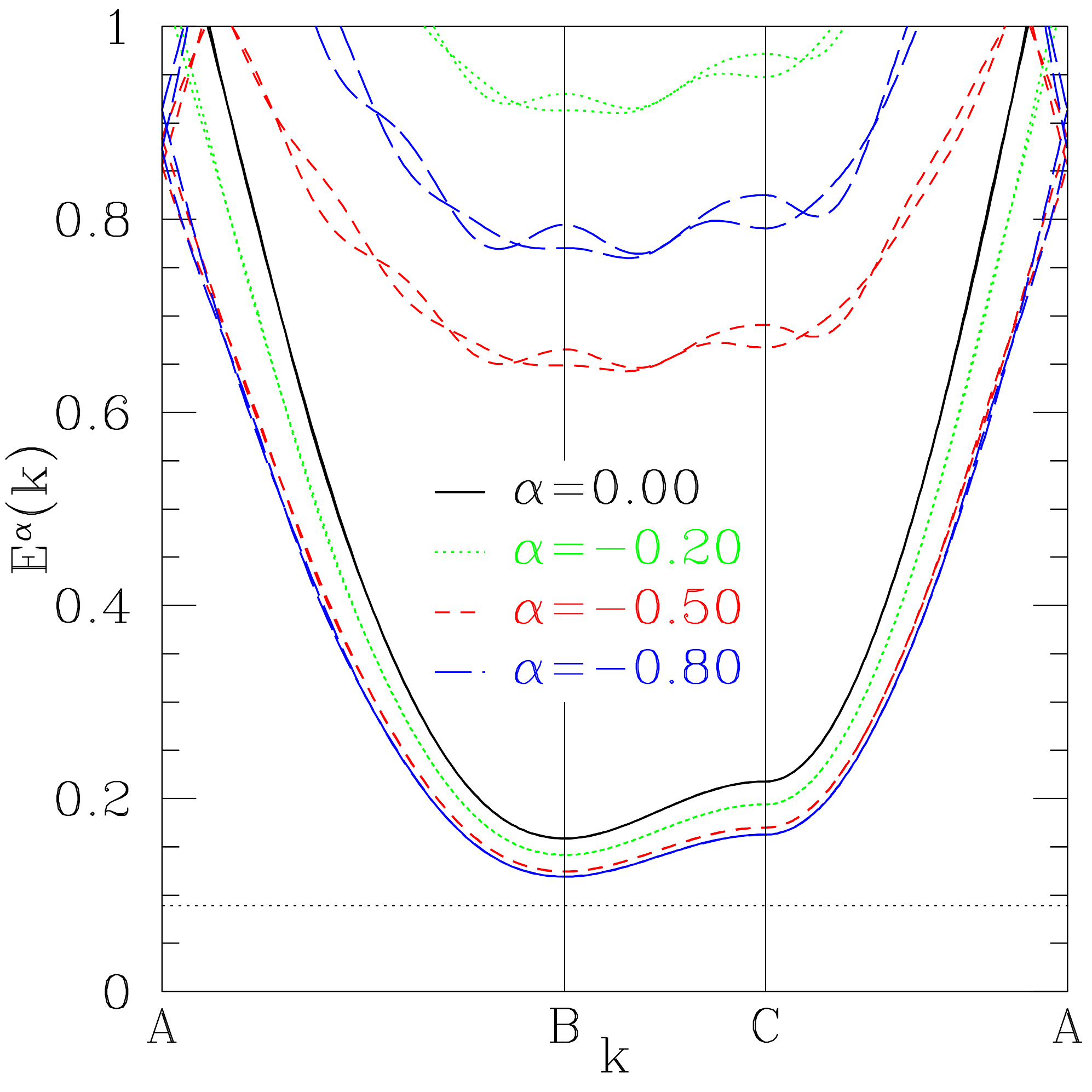}
 \caption{
(Color online) Variational vison (Eq.~\ref{eq:dressed_vison}) dispersion relation for different values
of $\alpha$,
along the path  A$\to$B$\to$C$\to$A in the Brillouin zone (see Fig.~\ref{fig:bril}).
The curve for $\alpha=0$ corresponds to the variational energies of
point-vison states (Eq.~\ref{eq:point_vison}).
The absolute minimum is $E_{\rm min}=0.119$, found at $(k_x,k_y)=(\pi/6,\pi/2)=B$
for a value $\simeq -0.8$ of the variational parameter $\alpha$.
The exact value
of the gap
(0.089)\cite{ivanov04} is marked as a dotted horizontal line.
In this calculation, the matrix elements $S_{ij}^\alpha$ are neglected for triangles $i$ and $j$
at distance $d\geq d_{\rm max}=10$.
This value is large enough to ensure
a perfect convergence of the spectrum below $E\lesssim 0.6$,
as can be checked from the fact that the dispersion curves obtained
with a smaller truncation distance ($d_{\rm max}\simeq 8.47$ -- shown here with the same colors)
are practically identical.
Some higher energy states, however, are not fully converged.}
 \label{fig:var_disp}
\end{figure}

Although the matrix elements of $\mathcal{H}$ are simple, the band structure is, however, not that of a simple tight-binding Hamiltonian, because of the non-orthogonality of the present variational vison states.
In Fourier space, $S^0(k)$ and $H^0(k)$ are $4\times4$ matrices.
The spectrum is obtained by solving the generalized eigenvalue problem
$H^0(k)\psi_k -E(k)S^0(k)\psi_k=0$.
The results are shown in Fig.~\ref{fig:var_disp} ($\alpha=0$ curve).
The minimum of $\epsilon_k$ is found at $(k_x,k_y)=(\pi/6,\pi/2)$, in agreement
with the Monte Carlo results.\cite{ivanov04,ralko06} The corresponding energy is $\epsilon_{\rm min}=0.16$,
which is significantly lower than the energy (0.56) found above for a completely localized point vison $|i\rangle$.
The bottom of this variational point vison band is, however, still high compared to Ivanov's\cite{ivanov04} estimate (0.089) of the gap in the vison sector. The next section provides
an improved family of variational states.

\subsection{Dressed vison}

The vison wave function
of Eq.~\ref{eq:point_vison}, differs from the RK wave functions only through minus signs. It
has the ``accidental'' property that dimer-dimer
correlations are the same as in the ground state.\footnote{In the kagome-lattice model of Ref.~\onlinecite{msp02},
this is an exact eigenstate.} 
As a consequence, the excitation energy of such a state is carried only by the kinetic part of the Hamiltonian.
Clearly, some {\it local} reweighting of the dimer configurations
in the vicinity of the vortex core would allow an optimized balance between the potential and kinetic costs
of the excitation, and would lead to an improved variational wave function.
This amounts to ``dressing'' locally the initial point-vison state by some
{\it even} -- but  fluctuating -- number of additional point visons.

As a simple improvement of the vison wave function, we introduce
a variational parameter $\alpha$
to reweight the configurations depending on their ``flipability'' at the core of the
vison. This gives the following vison state:
\begin{eqnarray}
 | i,\alpha \rangle& =&\frac{1}{\sqrt{\mathcal N}} \sum_c (-1)^{\hat N(i)}
        \left( 1 +\alpha  \hat F_i \right)
        | c \rangle \label{eq:dressed_vison},\\
        \hat F_i&=&\hat F_{r_1(i)}+\hat F_{r_2(i)} +\hat F_{r_3(i)}, \nonumber
\end{eqnarray}
where $\hat F_i|c\rangle=|c\rangle$ if the dimer configuration $c$ is flippable
around one of the three rhombi $r_1(i)$, $r_2(i)$, and $r_3(i)$
containing the triangle $i$, and $\hat F_i|c\rangle=0$ otherwise.


As for the point vison of Eq.~\ref{eq:point_vison}, the vison states of Eq.~\ref{eq:dressed_vison}
are not orthogonal and we have to evaluate their overlaps:
\begin{eqnarray}
    S^\alpha_{ij}=\langle i,\alpha | j,\alpha\rangle
    \label{eq:Sija}
\end{eqnarray}
Repeating the transformation leading to Eq.~\ref{eq:Saverage}, we get
\begin{equation}
    S_{ij}^\alpha=\epsilon_{ij} \left< (-1)^{\hat N(i,j)}
        \left( 1 +\alpha  \hat F_i\right)\left( 1 +\alpha  \hat F_j\right)
        \right>,
\end{equation}
which is an expectation value for a diagonal operator that we evaluate using Pfaffians.
In addition to the sign changes due to $(-1)^{\hat N(i,j)}$,
some entries of the Kasteleyn matrix have to be modified to incorporate
the flippability operators $\hat F_r$.
More precisely, counting only the coverings which satisfy
$\hat F_r=1$ is done by ``isolating'' the rhombus $r$, that is, by switching to zero in the Kasteleyn matrix the 14 bonds
which connect the sites of rhombus $r$ to their neighbors outside $r$.

Beyond some distance between the vison cores $i$ and $j$ (fourth neighbor on the hexagonal lattice),
no rhombus can touch simultaneously both triangles. In that case, it can be shown that
$S^\alpha_{ij}=S^0_{ij}$ is independent of $\alpha$.
The eigenvalues of the overlap matrix $S^\alpha(k)$ are
displayed in Fig.~\ref{fig:sk} for a few selected values of $\alpha$.
Although a full convergence as a function of the truncation distance $d_{\rm max}$ 
(see caption of Fig.~\ref{fig:sk}) has not been obtained, we believe that there is a finite gap for all
the values of $\alpha$ shown here, and that the dressed vison states are linearly independent.

The evaluation of the Hamiltonian matrix elements
\begin{eqnarray}
    H^\alpha_{ij}=\langle i,\alpha |\mathcal{H} |j,\alpha\rangle    
\end{eqnarray}
for dressed vison can still be done using Pfaffians, but
the algebraic manipulations are slightly more lengthy than the previous ones
and we refer the reader to Appendix~\ref{app:Hija}.
We find nonzero matrix elements $H_{ij}^\alpha$ up to (and including) the 9$^{\rm th}$ neighbor
(compared to first neighbor for point vison).


The results for the variational dispersion relation
are shown Fig.~\ref{fig:var_disp}.
The qualitative shape of the lowest band is almost unchanged compared to the point
vison states ($\alpha=0$), except for an almost uniform shift which lowers the gap.
The minimum of $E_k$ is found at $(k_x,k_y)=(\pi/6,\pi/2)=B$
for a variational parameter $\alpha\simeq-0.8$,
and the corresponding energy is $E_{\rm min}=0.119$,
about $33\%$ higher than the exact value.

The fact that simple wave functions like those of Eq.~\ref{eq:point_vison}
or Eq.~\ref{eq:dressed_vison} reproduce qualitatively the shape of the exact dispersion relation
is presumably due to the fact that the dimer-dimer correlation length is rather small at the RK point
(of order of one lattice spacing\cite{ms01b}).
As a consequence, the exact vison states only differ from Eq.~\ref{eq:point_vison}
at short distances from the vortex core, and the long-distance part (string), responsible
for the flux $\pi$ per hexagon, is essentially exact. This is, of course, no longer true
away from the RK point and in the direction of the crystal, where the correlation length rapidly grows.

In an attempt to extend this variational approximation {\it away} from the RK point, we computed
the gap of the dressed vison  states in perturbation theory to first order in $(1-V)$.
However, at this order, the gap turns out to close very slowly and does not lead to a meaningful estimate
for the critical $V$ at the liquid/crystal transition. This failure is closely related
to the remark above: the size of the core of the vison presumably grows  rapidly  away from the RK point,
a feature which cannot be accounted for with the present variational states.

\section{Conclusions}

In this paper, we have developed two simple approaches to describe 
the vison excitations of the QDM on the triangular lattice. The first one is based
on a  soft-dimer version of the model, which exactly takes the form of 
a {$\Z$} gauge theory. We have shown that a semiclassical spin-wave approximation
to the spectrum of the dual Ising theory captures the important fact that the 
disappearance of the RVB liquid is due to a vison 
condensation.\footnote{
In the frustrated Ising model formulation of the dimer model,
 $\sigma^x$  measures the presence (-1) or absence (+1)
of a vison, and $\sigma^z$ creates or annihilates a vison (see Eq.~\ref{eq:Psigmaz} for instance).
The classical ground state turns out to have $\langle \sigma^z \rangle=0$
in the dimer liquid phase and $\langle \sigma^z \rangle\ne0$ in the crystal phase (Sec.~\ref{sssec:cpd}).
So, the vison creation/annihilation operator acquires a finite expectation value at the transition,
the usual signature of a particle condensation.
This is almost identical to the standard hard-core-boson $\leftrightarrow$ spin-$\frac{1}{2}$ correspondence.
If the particle density is represented by $\frac{1}{2}(1-\sigma^x)$ (as here for visons), magnetic long-range order in the $z$
(or  $y$) direction (for the spin variables) is equivalent to Bose condensation (off-diagonal long-range order in the Boson operators).
An important difference is, however, that the number of visons
is {\it not} conserved. Only their {\it parity} is conserved. Accordingly, the vison condensed phase spontaneously breaks a discrete ($\Z$)
gauge symmetry ($\sigma^z$, which is not gauge invariant, acquires a finite expectation value in the crystal phase), and not a continuous [$U(1)$] one. Consequently, the condensed phase is gapped and does not have a Goldstone mode. In the dimer model, the spectrum is, therefore, gapless only {\it at} the transition.
} 
It also reproduces the qualitative shape of the vison spectrum in the
disordered phase and, more importantly, at the
transition (linear spectrum at the correct points of the Brillouin zone), as well as the spatial 
pattern of the ordered crystalline state.

The second approach is a variational approximation to the vison wave functions
at the RK point of the (hard-core) QDM. It
reproduces semiquantitatively the vison dispersion relation,
and provides a simple picture for the vison wave functions which goes beyond 
the naive point-vison approximation.

Beyond the problem of vison dispersion and condensation, we expect 
these approaches to be useful for other problems related to
vison excitations in QDM, in particular, their mutual interaction and their interaction
with vacancies and mobile holes. This is left for future investigations.

\section*{Acknowledgements}
We are very grateful to F.~Becca, M.~Ferrero, D.~Ivanov, V.~Pasquier, and A.~Ralko for
numerous insightful discussions. FM acknowledges the financial support of the Swiss
National Fund, of MaNEP, and of the R\'egion Midi-Pyr\'en\'ees through its
Chaire d'Excellence Pierre de Fermat program, and thanks the IPhT (Saclay) and the
Laboratoire de Physique Th\'eorique (Toulouse) for their hospitality.

\appendix

\section{Fourth order effective Quantum Dimer Model}
\label{sec:4th}

The goal of this Appendix is to derive an effective Hamiltonian for the
model of section II.A defined by the Hamiltonian
\begin{equation}
H=H_J+H_\Gamma=- J \sum_l \tau^x_l  - \Gamma \sum_i \prod_{l(i)} \tau^z_{l(i)}  
\end{equation}
in the limit $\Gamma/J\ll 1$ to fourth order in $\Gamma/J$. For simplicity,
let us define the unperturbed Hamiltonian $H_0\equiv H_J$ and the
perturbation $V\equiv H_\Gamma$. Since the Hilbert space of the model is restricted 
by the constraint that the number of dimers starting from a given site is
odd, the ground state manifold of the unperturbed Hamiltonian for positive
$J$ consists of all states having exactly one dimer emanating from each 
site. Let us denote by $P_0$ the projector onto the Hilbert space generated
by these states, and by $S$ the resolvent defined by
\begin{equation}
S=-\frac{1-P_0}{H_0-E_0}.
\end{equation}
It is easy to check that $V$ changes the parity of the number of
dimers. Indeed, the term of $V$ acting on the triangle
$i$ transforms the states with 0 and 3 (resp. 1 and 2) dimers around
the triangle $i$ into each other. This implies that $P_0VP_0=0$, 
and more generally that the effective Hamiltonian only contains even
powers of the perturbation. Thanks to the property $P_0VP_0=0$, the fourth
order contribution reduces to 3 terms, and the effective Hamiltonian up to
fourth order reads:
\begin{eqnarray}
H_{\rm eff} &=& P_0VSVP_0 + P_0VSVSVSVP_0 \nonumber \\
&-& \frac{1}{2}P_0\left(VS^2VP_0VSV + VSVP_0VS^2V\right)P_0 \nonumber
\end{eqnarray}

\begin{figure}
 \includegraphics[width=5cm]{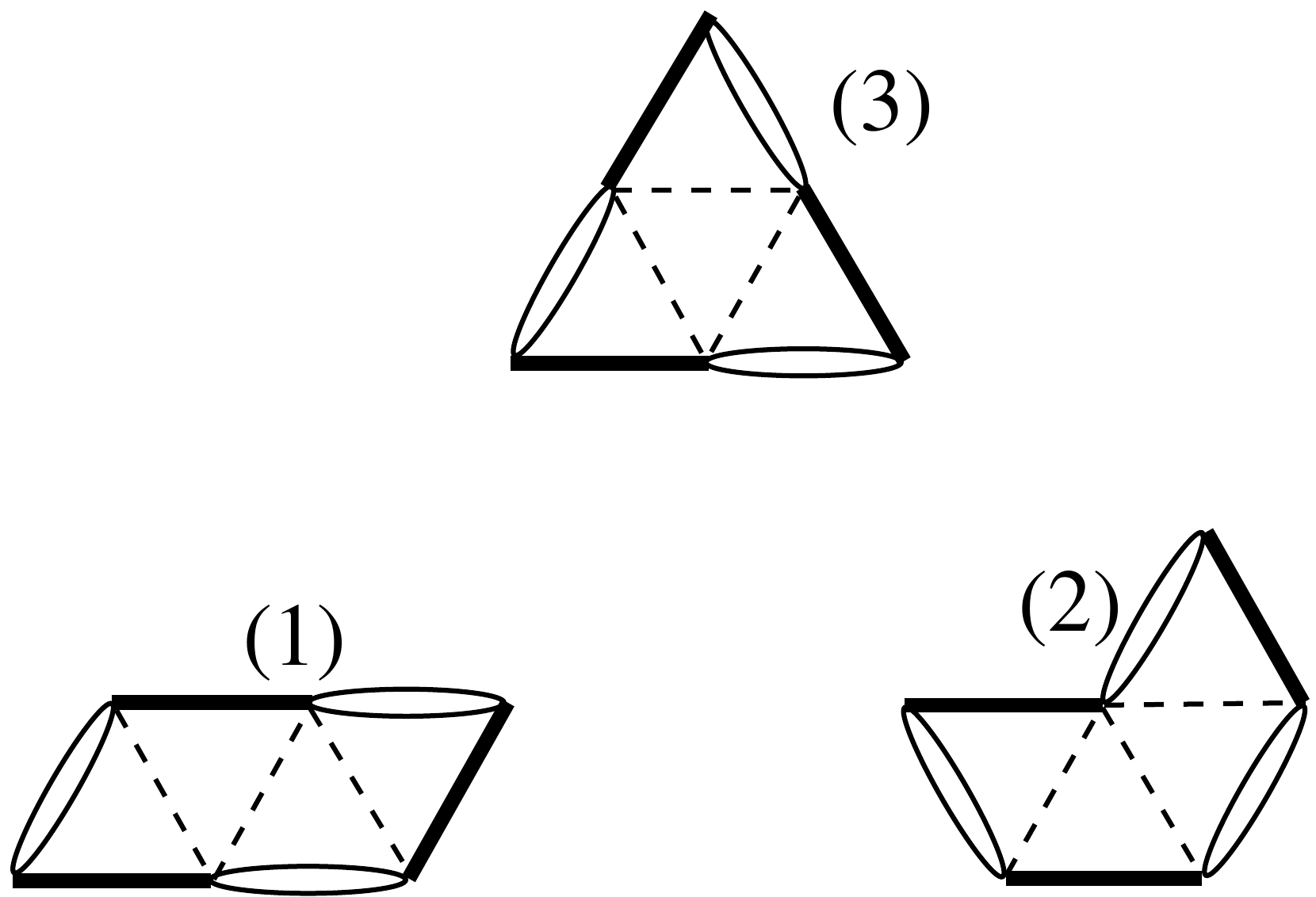}
 \caption{The three types of 6-site plaquettes around which dimer shifts are
 generated at fourth order in $\Gamma/J$. 
 }
 \label{fig:boucles6}
\end{figure}

Up to a constant, this effective Hamiltonian can be written as a QDM acting
on 4 and 6-site plaquettes. The 4-site Hamiltonian has the form of the
regular RK model with $t=\Gamma^2/J-\Gamma^4/J^3$ and $V=\Gamma^4/2J^3$.
The 6-site Hamiltonian only consists of kinetic terms that flip the dimers
around the three possible types of 6-site plaquettes shown in
Fig.~\ref{fig:boucles6}, with amplitude $-3\Gamma^4/4J^3$ for types (1) and (2),
and with amplitude $-\Gamma^4/J^3$ for type (3). 

\section{Hopping amplitude for the dressed visons}
\label{app:Hija}
For the dressed vison states (Eq.~\ref{eq:dressed_vison}), the equivalent of Eq.~\ref{eq:Hij0}
is
\begin{eqnarray}
    H_{ij}^\alpha&=&\frac{2}{\mathcal N} \sum_{r_0} \sum_{\begin{array}{c}(c,\bar c)\\F_{r_0}(c)=1\end{array}}  
        (-1)^{N(c,i)+ N(c,j)}\nonumber \\
        &&
        \times\left[ \langle c| (1+\alpha\hat F_i)+ \delta_{i,{r_0}}\langle \bar c|(1+\alpha\hat F_i) \right]\nonumber \\
        &&\hat\Pi_{r_0} 
        \left[ (1+\alpha\hat F_j)| c \rangle+ \delta_{j,{r_0}} (1+\alpha\hat F_j)| \bar c \rangle \right]
\label{eq:Hija}
\end{eqnarray}
Using the explicit form  of the projector $\hat\Pi_r$, we get
\begin{eqnarray}
\langle i,\alpha | \hat\Pi_{r_0} |j,\alpha \rangle&&=
\frac{1}{2\mathcal N} \sum_{\begin{array}{c}(c,\bar c)\\F_{r_0}(c)=1\end{array}}
(-1)^{N(c,i)+ N(c,j)} \nonumber \\
&&\times\left(1+\alpha F_i(c)-\delta_{i,{r_0}}(1+\alpha F_i(\bar c))\right)\nonumber \\
&&\times\left(1+\alpha F_j(c)-\delta_{j,{r_0}}(1+\alpha F_j(\bar c))\right)
\end{eqnarray}
Unlike Eq.~\ref{eq:Hij0b},
both the coverings $c$ and $\bar c$ enter the expression.
This does not have the form of a diagonal observable and the terms $F_{r}(\bar c)$ (with $r\in\left\{r_1(i),r_2(i),r_3(i),r_1(j),r_2(j),r_3(j)\right\}$)
need to be eliminated
to allow for an evaluation with Pfaffians.
By inspecting the possible relative positions of two rhombi $r_0$ and $r$, one arrives
at the following two relations:
\begin{itemize}
 \item If two rhombi $r_0$ and $r$ have 0,1,3 or 4 sites in common,
 $F_{r}(\bar c)=F_{r}(c)$ for any pair of configurations
 $(c,\bar c)$ which differ by a flip around the rhombus $r_0$.
 \item It $r_0$ and $r$ have 2 sites in common, let us call $b$ the bond of
 $r$ which does not touch $r_0$. Then we have
 $F_{r}(\bar c)=D_b(c)(1-F_{r}(c))$, where $D_b(c)=1$ if $b$ is occupied
 by a dimer of $c$, and 0 otherwise.
\end{itemize}
We may combine the two cases above into some compact notation $F_r(\bar c)= A_{r,r_0}(c)$,
valid for any pair of configurations
 $(c,\bar c)$ which differ by a flip around the rhombus $r_0$.
Accordingly, we define an operator $\hat A_{i,r_0}=\hat A_{r_1(i),r_0}+\hat A_{r_2(i),r_0}+\hat A_{r_3(i),r_0}$ for each triangle $i$ and rhombus $r_0$.
The matrix element of Eq.~\ref{eq:Hija} is now expressed as the expectation value of a
diagonal operator:
\begin{eqnarray}
 H_{ij}^\alpha&=&\frac{\epsilon_{ij}}{2}\sum_{r_0}
 \left<
(-1)^{\hat N(i,j)}\hat F_{r_0} \right.\nonumber\\
&&\times
 \left(
    1+\alpha \hat F_i-\delta_{i,{r_0}}(1+\alpha A_{i,r_0})
 \right)\nonumber\\
&&\times\left. \left(
    1+\alpha \hat F_j-\delta_{j,{r_0}}(1+\alpha A_{j,r_0})
    \right)
 \right>
\end{eqnarray}
This expression has to be expanded into a polynomial
in $\hat F$ and $\hat D$ operators 
before each term can be evaluated thanks to the Pfaffian of an appropriate
Kasteleyn matrix. As before the $(-1)^{\hat N(i,j)}$ introduces some sign changes, and each $\hat F$ (or $\hat D$) operator
requires to ``isolating'' the corresponding rhombus (or bond) by switching to zero the corresponding matrix elements. Several tens of terms typically appear for each pair of triangles $(ij)$, and an automated treatment by computer had to be coded to obtain the hopping amplitudes. The results of Fig.~\ref{fig:var_disp} represent
several hundreds of CPU hours using the software
Maple to generate all the correlators and evaluate the corresponding Pfaffians on a finite lattice with
$28\times28$ sites.


\begin{thebibliography}{99}

\bibitem{ms01b}
R. Moessner and S.~L. Sondhi,
\journal{prl}{86}{1881}{2001}.

\bibitem{msp02}
G. Misguich, D. Serban, and V. Pasquier,
\journal{prl}{89}{137202}{2002}.

\bibitem{ms03}
R. Moessner and S.~L. Sondhi,
\journal{prb}{68}{054405}{2003}.

\bibitem{rk88}
D. S. Rokhsar and S. A. Kivelson,
\journal{prl}{61}{2376}{1988}.

\bibitem{ioselevich02}
A. Ioselevich, D. A. Ivanov, and M. V. Feigelman,
\journal{prb}{66}{174405}{2002}.

\bibitem{ivanov04}
D. Ivanov,
\journal{prb}{70}{094430}{2004}.

\bibitem{ralko05}
A. Ralko, M. Ferrero, F. Becca, D. Ivanov, and F. Mila 
\journal{prb}{71}{224109}{2005}.

\bibitem{lca07}
A. Laeuchli, S. Capponi
and F.~F.~Assaad,
\href{http://dx.doi.org/10.1088/1742-5468/2008/01/P01010}{J. Stat. Mech., P01010 (2008)}.

\bibitem{fmo06}
S.~Furukawa, G.~Misguich and M.~Oshikawa,
\href{http://dx.doi.org/10.1088/0953-8984/19/14/145212}{J. Phys.: Condens. Matter {\bf 19}, 145212 (2007)}.

\bibitem{fm07}
S.~Furukawa and  G.~Misguich,
\journal{prb}{75}{214407}{2007}.

\bibitem{js91}
R. A. Jalabert and S. Sachdev,
\journal{prb}{44}{686}{1991}.

\bibitem{sv00}
S. Sachdev and M. Voijta,
\journal{J. Phys. Soc. Jpn}{69}{Suppl. B 1}{2000};
\journal{cond-mat}{9910231}{0}{0}.

\bibitem{sf00}
T. Senthil and M.~P.~A.~Fisher,
\journal{prb}{62}{7850}{2000}.

\bibitem{sf01}
T. Senthil and M. P. A. Fisher,
\journal{prl}{86}{292}{2001};
\journal{prb}{63}{134521}{2001}.

\bibitem{wegner71}
F. Wegner,
\href{http://dx.doi.org/10.1063/1.1665530}{J. Math. Phys. {\bf 12}, 2259 (1971)}.

\bibitem{msc00}
R. Moessner, S. L. Sondhi, and P. Chandra,
\journal{prl}{84}{4457}{2000}.


\bibitem{ms01}
R. Moessner and S. L. Sondhi
\journal{prb}{63}{224401}{2001}.

\bibitem{ralko06}
A. Ralko, M. Ferrero, F. Becca, D. Ivanov and F. Mila,
\journal{prb}{74}{134301}{2006}.

\bibitem{ralko07}
A. Ralko, M. Ferrero, F. Becca, D. Ivanov and F.~Mila,
\journal{prb}{76}{140404(R)}{2007}.


\bibitem{msf02}
R. Moessner, S.~L.~Sondhi, and E.~Fradkin,
\journal{prb}{65}{024504}{2002}.


\bibitem{villain77}
J. Villain,
\href{http://www.iop.org/EJ/abstract/0022-3719/10/10/014/}{J. Phys C {\bf 10}, 1717 (1977)}.

\bibitem{rc89}
N. Read and B. Chakraborty,
\journal{prb}{40}{7133}{1989}.

\bibitem{k61}
P. W. Kasteleyn,
\href{http://dx.doi.org/10.1016/0031-8914(61)90063-5}{Physica {\bf 27}, 1209 (1961)}.


\bibitem{krauth}
W. Krauth, {\it Statistical Mechanics: Algorithms and Computations}, Oxford University Press, 2006.


\bibitem{wen02}
X.-G. Wen,
\journal{prb}{65}{165113}{2002}.




\end{thebibliography}
\end{document}